\documentclass[fleqn,usenatbib]{mnras}

\usepackage{newtxtext,newtxmath}

\usepackage[T1]{fontenc}

\DeclareRobustCommand{\VAN}[3]{#2}
\let\VANthebibliography\thebibliography
\def\thebibliography{\DeclareRobustCommand{\VAN}[3]{##3}\VANthebibliography}

\usepackage{graphicx}	
\usepackage{amsmath}	
\usepackage[czech]{babel}
\usepackage{float}
\usepackage{multirow}
\usepackage{makecell}

\usepackage{xcolor,colortbl}

\definecolor{Gray}{gray}{0.85}
\definecolor{LightCyan}{rgb}{0.88,1,1}

\makeatletter
\newcommand{\thickhline}{%
    \noalign {\ifnum 0=`}\fi \hrule height 1pt
    \futurelet \reserved@a \@xhline
}
\newcolumntype{"}{@{\hskip\tabcolsep\vrule width 1pt\hskip\tabcolsep}}
\makeatother

\title[SED proﬁles from AGN...]{Spectral Energy Distribution profiles from AGN accretion disc in multi-gap setup}

\author[M. \v{S}tolc et al.]{
Marcel \v{S}tolc,$^{1,2}$\thanks{Corresponding author. E-mail: stolcml@gmail.com}
Michal Zaja\v{c}ek,$^{3}$
Bożena Czerny,$^{4}$ \&
Vladimír Karas $^{2}$
\\
$^{1}$Astronomical Institute, Faculty of Mathematics and Physics, Charles University, V Hole\v{s}ovi\v{c}k\'ach 2, 180\,00 Prague, Czech Republic\\
$^{2}$Astronomical Institute of the Czech Academy of Sciences, Bo\v{c}n\'{\i} II 1401, 141\,00 Prague, Czech Republic\\
$^{3}$Department of Theoretical Physics and Astrophysics, Faculty of Science, Masaryk University, Kotl\'a\v{r}sk\'a 2, Brno, 611\,37, Czech Republic\\
$^{4}$Center for Theoretical Physics, Polish Academy of Sciences, Al. Lotnik\'ow 32/46, 02-668 Warsaw, Poland
}

\date{Accepted 2023 April 3. Received 2023 April 3; in original form 2022 August 3}

\pubyear{????}

\begin{document}
\label{firstpage}
\pagerange{\pageref{firstpage}--\pageref{lastpage}}
\maketitle

\begin{abstract}
\textcolor{black}{ Spectral Energy Distribution (SED) of the broad-band continuum emission from black-hole accretion discs can serve as a tool to measure parameters of the central body and constrain the geometry of the inner accretion flow. We focus on the case of an active galactic nucleus (AGN), with an accretion disc dominating the UV/optical bands. We parameterize the changes in the thermal and power-law components, which can reveal the diminution of the emissivity. To this end we explore the effects of gaps in the accretion disc and the emerging SED that can be caused by the presence of either (i) the inner, optically thin, radiatively inefficient hot flow;
(ii) a secondary black hole embedded within the accretion disc; or (iii) a combination of both components. We suggest that the resulting changes in the SED of the underlying continuum can help us to understand some departures from the standard-disc scenario. 
We estimate that the data required for such a project must be sampled in detail over the far-UV to soft X-ray bands during the interval of about a month corresponding to the characteristic variability timescale of an AGN. Detecting a gap at intermediate radii of a few 100 gravitational radii would require quality photometry with uncertainties up to $\sim1\%$. The presence of the central cavity in the standard disc can be recovered in UV photometric data with an accuracy of 5\% and better. We show the effect of the intrinsic reddening of the source 
and demonstrate when it can be disentangled.}
\end{abstract}




\section{Introduction}

Supermassive black holes (SMBH) are located in the centres of majority of galaxies \citep{1982MNRAS.200..115S,magorian}. Within the framework of General Relativity, SMBHs in nuclei of galaxies are characterized by just two parameters: mass $M_\bullet$ encompassed within the event horizon, and the associated angular momentum (spin parameter), $a\equiv J/Mc^2$ \citep[cf. No Hair Theorem in][]{misner}. SMBHs reside in the heart of the central system, which is generally characterized by a quasispherical stellar system or a nuclear star cluster as well as circumnuclear gaseous-dusty rings and filaments \citep[see][for a review]{2021bhns.confE...1K}. The SMBH mass is closely correlated with the large-scale properties of the host galaxy, including the velocity dispersion, bulge mass, or the luminosity, which implies the mutual coevolution of SMBHs and their host galaxies, most likely via the mechanism of AGN feeding and feedback \citep{2009ApJ...698..198G,2013ARA&A..51..511K,2022NatAs...6.1008Z}.   Naturally, there are various assumptions that need to be verified. In particular, as the first approximation, we assume the black hole to be in a steady state, not disturbed by the surrounding nearby bodies \citep{1973CMaPh..31..161B,1987thyg.book.....H}, and its electric charge is set to zero as it is typically negligible \citep{1972PhRvD...6.1476W,2018MNRAS.480.4408Z}. 

The methodology to study the temporal evolution of thin accretion discs has been first developed in the Newtonian regime \citep{1974MNRAS.168..603L}.
It was shown that the inner accretion disc structure is sensitive to the viscosity law assumed, and the standard $\alpha$-model discs become unstable to clumping in their inner regions, where the radiation pressure dominates over the gas pressure  \citep[e.g.][]{1975ApJ...200..187E}. The inner solution emerges, where the medium is optically thin and radiatively inefficient, so that it defines an inner edge to the outer (standard) solution. 
The General Relativistic disc evolution has been introduced by \citet{1974ApJ...191..499P}, and its various aspects have been explored in seminal works by \citet{2000ApJ...529..119R}, \citet{2018MNRAS.481.3348B}, and further references cited therein.    

 The accreted plasma possesses angular momentum as it moves around the central body and it exhibits the magneto-rotational instability (MRI), which determines the viscous mass transport \citep{balbus_MRI}. Based on the intrinsic properties of the gas and the boundary conditions of the system, the emerging structure can be modelled in terms of a geometrically thin accretion disc \citep{shakura}, a slim accretion disc \citep{abramowicz88}, or a geometrically thick fluid torus \citep{fishborne76}.

The standard Shakura-Sunyaev scenario has been particularly successful in explaining the thermal component of the spectral energy distribution (SED) of active galactic nuclei, including the ``Big Blue Bump'' \citep[see e.g.,][and further references cited therein]{1987ApJ...321..305C,2015MNRAS.446.3427C}. However, various perturbations can cause deviations from the standard-disc scheme. They can be caused, for instance, by
\begin{itemize}
\item[(i)] gravitational perturbation caused by the presence of a second compact body, ranging from stellar-sized objects to massive black holes \citep{1991MNRAS.250..505S}; accumulation of more objects in disc ``migration traps'' has also been analyzed \citep{2016ApJ...819L..17B},
\item[(ii)] hydrodynamical effects due to large-scale inflows (Bondi-type accretion) and outflows/winds \citep{Ricci}, including also supernova explosions \citep{2021ApJ...906...15M}, 
\item[(iii)] the transition of an optically thick accretion flow into an optically thin, hot flow in the inner regions \citep{1995ApJ...438L..37A,1996PASJ...48...77H,1998PASJ...50..559K}.
\end{itemize}
The mentioned effects can lead to non-standard radial profiles of the physical characteristics of the accretion flow, which we describe as a collection of transient (co-planar) discs or tori with a gap, or a structure of nested rings and gaps \citep{2011MNRAS.418..276S,2015ApJS..221...25P,wds_stolc}.

Secular or stochastic perturbations of stellar motion can lead to the scenario, in which a field star becomes gravitationally bound to the central body \citep{1991MNRAS.250..505S,2017ARA&A..55...17A}. The length-scale where the stellar dynamics is dominated by the central black hole is given by the radius of gravitational influence \citep[e.g.][]{2013degn.book.....M},
\begin{equation}
    R_{\rm inf} = \frac{GM_{\bullet}}{\sigma_{\star}^2}\,
    \simeq 48  \left(\frac{M_{\bullet}}{10^9\,M_{\odot}}\right) \left(\frac{\sigma_{\star}}{300{\rm km\,s^{-1}}}\right)^{-2}\,{\rm pc}\,,
    \label{eq_influence_radius}
\end{equation}
where $\sigma_{\star}$ is the one-dimensional velocity dispersion of stars in the bulge estimated using $M_{\bullet}$-$\sigma_{\star}$ relation \citep{2009ApJ...698..198G}. Eq.~\eqref{eq_influence_radius} expresses the radius of the sphere where the stellar cluster of $\sim 2\times 10^9\,M_{\odot}$ could in principle be bound to the SMBH.
The initial parameters of the perturbing star, such as non-zero inclination and the eccentricity of the trajectory, change as a result of orbital energy dissipation and repetitive star-disc interactions \citep{1991MNRAS.250..505S,subr,karas}. 
The length-scale of the standard thin accretion disc in the AGN is several orders of magnitude smaller than the gravitational influence radius of the SMBH. This is first given by the ability of the captured flow to circularize at a certain radius, and second, the outer radius of the stationary disc is given by the gravitational stability condition \citep{2004MNRAS.354.1177S}. 

Once in the orbital plane, given the ratio of the accretion disc scale-height and the star, a gap can be formed. Whether a gap is created or filled, the secondary body will spiral down to the centre due to gravitational emission. Eventually the orbiter reaches the tidal radius and becomes either totally or partially disrupted \citep{Hills,Rees}. The resulting gaseous trace can in principle be revealed as a spectral feature \citep{2020SSRv..216...85S,2019AN....340..570S}.
An alternative mechanism of the gap formation can be an outcome of late-phase galactic mergers, when the secondary black hole
orbiting around the primary one clears a trail in the accretion disc that can be seen in the spectral energy distribution \citep[e.g.][]{gultekin}.
Let us note that the drop in the accretion rate takes place in the inner parts of the accretion disc and it leads to a decrease in the optical depth of the material, and rapid rise of the inflow velocity as well as the temperature \citep{ichimaru1977,yuan_naryan_2014}. The cooling becomes dominated by advection, and the hot, diluted plasma acts effectively as an inner gap in the optically thick thermal emission. 

Furthermore, \cite{Ricci} state that the gap formation due to the tidal disruption stream may explain some of the observed ``Changing-Look'' events with characteristic changes in the X-ray properties of the corona. The reason for such a behaviour is the depletion of accretion flow in the inner region of the accretion disc, to be precise from tidal radius downwards, which can lead to the X-ray cut off \citep{Ricci}. Changing-Look events may play a role in shaping the emergent flux and the spectral line profiles induced by the continuum variability \citep{ngc1566_gap_2022}.  

In this paper, we present a study of SED thermal signatures caused by the effect of 
\begin{itemize}
    \item ADAF in the inner part of the disc (model A),
    \item secondary IMBH or SMBH located further away (model B),
    \item both cases A and B combined (model C).
\end{itemize}

The structure of the article is as follows. In Section~\ref{sec_model_outline}, we describe our model in general as a multiple-gap system, whereas in its subsequent subsections we formulate in detail the physical conditions underlying the proposed models (having at the most two gaps). Section~\ref{sec_results} contains the justification of our assumptions and describes the main results of the study. In Section~\ref{sec_discussion}, we discuss our results and possible future prospects of the scenario. We summarize our results in Section~\ref{sec_conclusions}.

\begin{figure}
    \centering
    \includegraphics[width=\columnwidth]{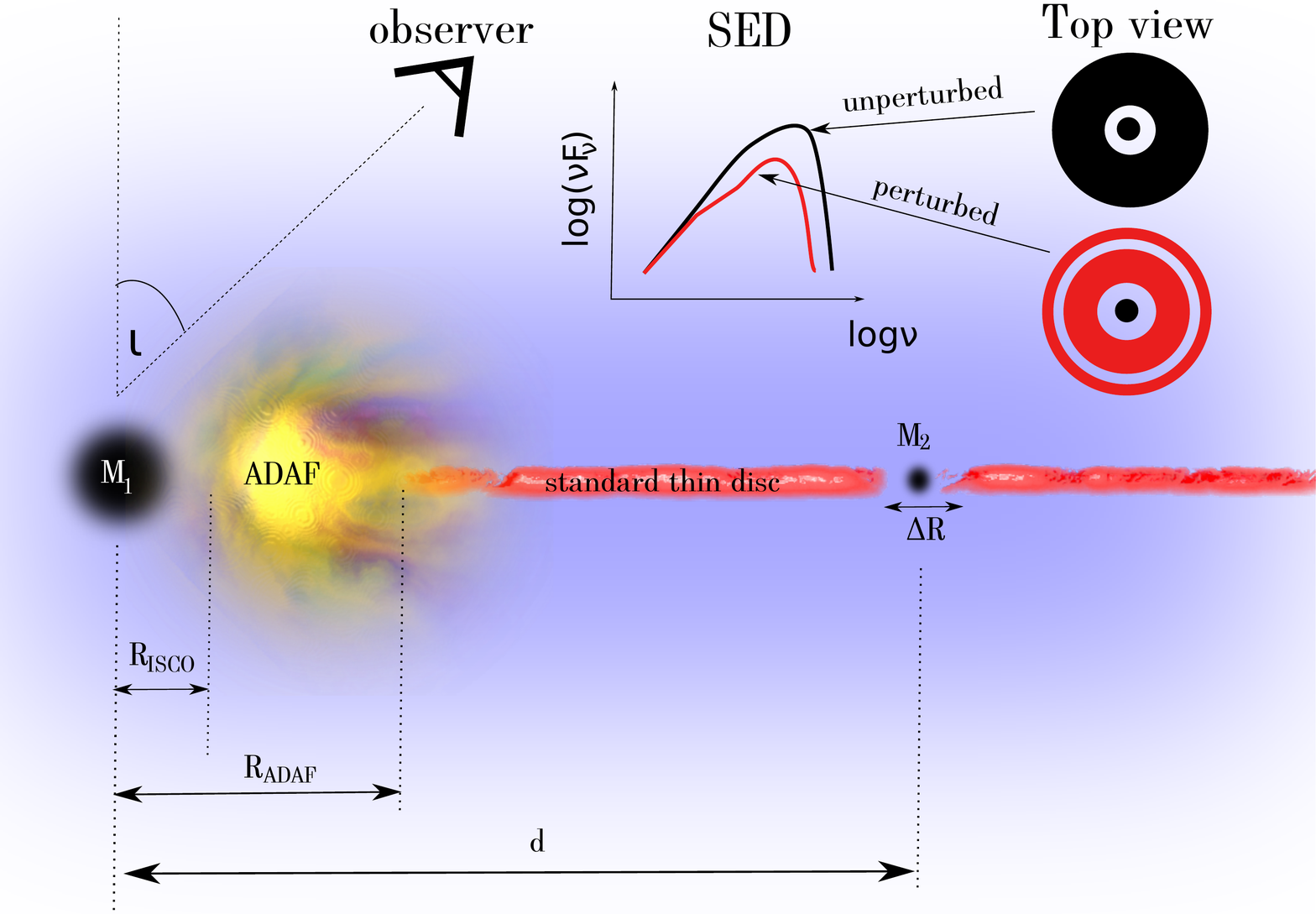}
    
    \caption{Sketch of the model components shown in an azimuthal section: an equatorial standard accretion disc (red), a central SMBH (a large black circle), a compact inner ADAF component (yellow), and a secondary, less massive black hole orbiting in the disc plane (smaller black circle). The hot plasma present in the ADAF region extends inside the standard accretion disc down to the ISCO, where the flow is ionized; it effectively creates a central gap in the thermal emission. A secondary black hole is placed further out, where it gives rise to a second gap in the standard disc, whose size depends on the mass of the orbiting perturber as well as its distance. Parameters of these components are explored in different flavours of three models A--C. Effects of curved space-time are described in terms of non-rotating (Schwarzschild) metric. The top right figure inset depicts the impact of the studied perturbations of the standard thin disc on the UV/optical continuum spectral energy distribution; see the text for details (figure components not to the scale).}
    \label{fig:disc_gap_ADAF.pdf}
\end{figure}

\section{Model outline and analytical estimates}
\label{sec_model_outline}

The basic assumption is the axial symmetry and the quasistationarity of the system. The main element of the system is the disc surrounding the central black hole that can be approximated by the standard Shakura-Sunyaev \citep{shakura} accretion disc.
The central black hole has a constant mass, the value of its charge is zero as well as that of its spin, i.e. we consider a Schwarzschild black hole. Furthermore, we do not take into account the effects of magnetic field in the system other than the initiation of the MRI \citep{balbus_MRI} needed for the accretion to proceed. 

The spectral energy distribution of the disc consists of a set of black-body temperature profiles, for which the temperature is a function of radius \citep{Frank},
\begin{equation}
    T(R) = \left(
    \frac{3GM_{\bullet}\dot{M}}{8 \pi \sigma R^3} f_{\rm{R}} \right)^{1/4},
    \label{eq_temperature}
\end{equation}
 where $f_{\rm{R}}$ includes the relativistic corrections and the impact of disc truncation \citep{2008bhad.book.....K}. In the Newtonian (non-relativistic) approach, assuming the prescription for a standard Shakura-Sunyaev thin accretion disc, the correction factor reads as $f_{\rm R}=1-\sqrt{R_{\rm{in}}/ R}$. Eq.~\eqref{eq_temperature} implies the characteristic temperature profile of the disc material, $T(R)\approx T_{\bullet} (R/R_{\rm ISCO})^{-3/4}$. The characteristic temperature $T_{\bullet}$ evaluated at the innermost stable circular orbit (ISCO) of the non-rotating black hole, $R_{\rm ISCO}=6R_{\rm g}$, increases for the decreasing black hole mass as $T_{\bullet}\propto M_{\bullet}^{-1/4}$, i.e. for $M_{\bullet}=10^9\,M_{\odot}$ we get $T_{\bullet}=130\,000\,{\rm K}$ for the Eddington ratio of $0.5$ and $12\%$ radiative efficiency, which falls into the UV domain with the peak wavelength of $\lambda\simeq 22\,{\rm nm}$.

In the further analysis, we represent the accretion disc as a system of infinitesimal mass rings and assume each of them radiating as a black body. By integrating over all of the rings we obtain the total luminosity \citep{Frank}
\begin{equation}
   \nu L_{\nu}(\nu) = \frac{16\pi^2 h \nu^4 \cos i}{f_{\rm{col}}^{4} c^2} \int ^{R_{\rm{out}}}_{R_{\rm{in}}} \frac{R\, dR}{e^x-1},
    \label{eq_luminosity}   
\end{equation}
where $x=\frac{h\nu}{k f_{\rm{col}}T(R)}$, $i$ is the viewing angle of the source (inclination), and $f_{\rm col}$ is the color correction factor that defines the spectral hardening with respect to the standard scenario \citep{2019ApJ...874...23D}. 

The latter expression represents the thermal component from the complete disc, extending from the inner to the outer radius, $R_{\rm in}$ and $R_{\rm out}$ \citep[e.g.][]{2016ApJ...832...22S}. Let us note that our main goal is to simulate the thermal continuum in the presence of multiple gaps, which reduces the signal originating from affected radii in eq. (\ref{eq_luminosity}). The total radiation of the perturbed disc with the system of $N$ gaps ($R_{\rm{gap-in}}, R_{\rm{gap-out}}$)  can then be expressed using the following formula  
\begin{equation}\label{gap_luminostiy}
     \nu L_{\nu}(\nu) = \frac{16\pi^2 h \nu^4 \cos i }{f_{\rm{col}}^{4}c^2}  \bigg ( \int^{R_{\rm{out}}}_{R_{\rm{in}}} 
     -\sum^{\rm{N}}_{i=1} \int^{R_{\rm{gap-\, out\,,i}}}_{R_{\rm{gap-in \,, i}}} \bigg) \frac{R\, dR}{e^x-1} ,
\end{equation}
where the relativistic effects are neglected (they are not of critical importance for the continuum originating at larger distances), and we also ignore any contribution to the total radiation from the gap.

\textcolor{black}{ Recently, \citet{2022ApJ...939L...2Z} considered a sequence of models with different location of the disc inner edge, the color correction factor, and values of different accretion rate. Let us note that although $f_{\rm col}$ influences the emerging SED primarily in the X-ray band, there is some residual effect extending to the UV and optical regime \citep{2012MNRAS.420.1848D}.}

\subsection{Inner gap (the central region)}

The standard accretion disc terminates at the ISCO \citep{1972ApJ...178..347B}. Indeed, this special location of the inner edge has been reported  in many observations of the inner accretion flow structure \citep[e.g.][]{tanaka95, Reynolds_2003}. However, a more realistic description of the accretion flow can lead to a significant displacement of the inner edge under various circumstances that invalidate some of the critical assumptions of the standard scenario \citep[e.g.][]{Krolik_2005}. Moreover, the very definition of the inner
termination radius can be understood in several different ways: the place where the inflow velocity starts growing and density starts dropping rapidly, where the emission properties change abruptly, etc. 

We adopt the scenario where a diluted, optically thin, highly ionized plasma fills the central region, so that it effectively forms an inner gap in the thermal emission as its radiative contribution to the SED in the UV spectral bands is negligible. The size of this component then determines the inner boundary of the standard accretion disc in our description. In fact, the advection-dominated part of the inflow serves as a working hypothesis to represent the adopted scenario, as indicated in Figure \ref{fig:disc_gap_ADAF.pdf}. Despite the ongoing analysis concerning the geometry, the size, and the very nature of the hot emitting plasma, frequently named an accretion disc corona, we can relate its structure to the toroidal (i.e. non-spherical) shape of the geometrically thick, rotating accretion inflow which develops inside the outer standard accretion disc. This helps us to model the effect of the central gap and discuss the influence of its parameters in specific terms.

\textcolor{black}{ We parameterize the inner gap by its radius:
\begin{equation}
R_{\rm in} = R_{\rm ADAF},
\end{equation}
and we set the no-torque boundary condition there. \citet{2022ApJ...939L...2Z} advocated the use of additional parameter setting the no-torque condition inside the ADAF flow. This indeed accounts for the possible viscous transfer of angular momentum by the hot flow, but such a new parameter cannot be easily constrained since the magnetic field action in the hot flow is complex, and vigorous fully ionized outflow can carry most of the angular momentum causing even the flow to be supersonic \citep[e.g. a series of models][well fitting the binary black hole systems]{ferreira2006,marcel2019}. So the torque may not act there, and we did not introduce any additional parameter to account for ADAF torque.}

     The ADAF in the inner parts of the accretion disc \citep{yuan_naryan_2014,1996ApJ...471..762A,1994ApJ...428L..13N} is thus essentially shifting the inner radius of the accretion disc { outwards}, otherwise set by default to $R_{\rm{ISCO}} = 6 R_{\rm{g}}$. 
     
\begin{figure}
    \centering
    \includegraphics[width=\columnwidth]{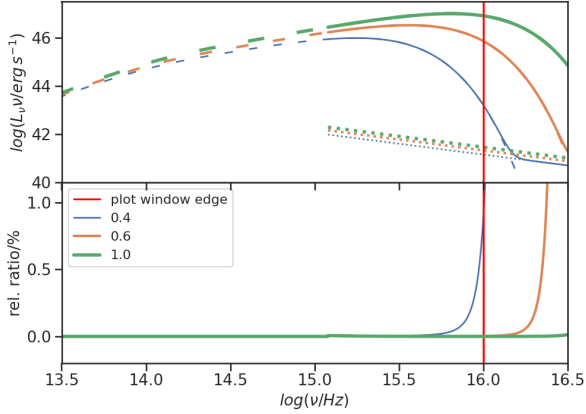}
    \caption{\textcolor{black}{ The examples of the broad band SED from the disc and the contribution from the hot ADAF plasma emission, scaled using Equation~\ref{uv_xray_rel} for varying values of $\dot{m}$\textcolor{black}{, with the powerlaw index $\Gamma$ set to 1.9 and the $E_{\rm min}$ calculated at $2500\, \mathring{A}$ and corresponding to $\approx 5\,{\rm eV}$.} \textcolor{black}{ The mass of the primary is set to $10^9M_{\odot}$.} The solid red vertical line marks the plot window edge of the respective Figures illustrating SEDs for models involving the ADAF component. Top panel: Dashed lines show SED without ADAF contribution. Solid lines show SED with an ADAF contribution. Finally dotted lines show only ADAF contribution. Bottom panel: Relative ratio between SED with and without ADAF contribution. }}
    \label{fig:power_law}
\end{figure}

     \textcolor{black}{  
     Apart from the cold disc emission, observations clearly show the presence of the X-ray power law component. This Comptonized emission comes from the medium defined as the hot corona, in our case the ADAF region forming the central cavity effect in the thermal component. This coronal emission is always present although it is relatively less important for the brightest sources. This trend is reflected in the correlation between the UV and X-ray luminosities \citep[e.g.][]{tananbaum1979,risaliti2015,lusso2016,2020A&A...642A.150L}. \citet{netzer} gives the following expression: 
\begin{equation}\label{uv_xray_rel}
    \log L_{\nu\,(2keV)} = 0.62\log L_{\nu \, (2500 \mathring{A})}+7.77.
\end{equation}}
\textcolor{black}{ This power law emission does not contribute much to the far UV, partially because it is a Compton scattering effect so the emission starts at the frequency of the scattered disc photon, and partially because of the normalization of this component which is low. \textcolor{black}{ We set the powerlaw index $\Gamma$ to 1.9 and $E_{\rm min}$ is calculated at $2500 \mathring{A}$ corresponding to $\approx 5 eV$.} We illustrate this effect in Figure~\ref{fig:power_law}. In further considerations, we neglect the ADAF contribution to the UV emission.}

     \textcolor{black}{ Apart from the formation of an inner ADAF, there may be also another possibility to create an inner gap in the standard disc.}
     It is interesting to note that recently the Changing-Look behaviour of some AGN on timescales of years was tentatively attributed to a sudden displacement of the inner radius \citep{ngc1566_gap_2022}, which can be attributed to a non-stationary behaviour of the inner disc. Also, the perturbation of the standard accretion disc due to an orbiting massive perturber (a secondary black hole) can produce gaps and modulate both the inflow and the outflow, which leads to Changing-Look luminosity of the source with a yearly variability \citep[e.g.][]{2020arXiv201202608S,2021ApJ...917...43S,sniegowska,sniegowska22}.
    
\subsection{Outer gap by a secondary companion} 
\label{sect:outer_gap}

    In our models, we attribute the formation of a second gap located at larger distances from the SMBH to the presence of a secondary compact object embedded in the accretion-disc plane. Once the object, for example a star or a black hole, which is trapped in the gravitational potential of the central black hole, reaches the orbital plane of the disc, one can expect the gap to be formed. The ability to form a stable gap depends on the sphere of influence of the particular perturbing body with respect to the disc scale-height \citep[e.g.][]{subr,karas}. When we neglect outflows emanating from the secondary body, the typical size of the gap is related to the tidal (Hill) radius. As we show later, a star cannot open a large gap in the disc, but a secondary black hole can do that. Considering a secondary black hole with the mass of  $M_2=10^{-2} M_1$, e.g. $M_{2}=10^7 M_{\odot}$ merging with the primary black hole of the mass $M_1=10^9 M_{\odot}$, a gap is created at the distance 
    $d$, and its size $\Delta R$ can be estimated as
    twice the Hill radius of the secondary component:
    \begin{align}
   \frac{\Delta R}{R_{\rm{g}}}   &\approx \frac{2d}{R_{\rm g}}\left(\frac{M_2}{3 M_1} \right)^{1/3}\,\notag\\
   &\approx 29.9 \left(\frac{d}{100R_{\rm g}}\right)\left(\frac{M_2}{10^7\,M_{\odot}}\right)^{1/3}\left(\frac{M_1}{10^9\,M_{\odot}}\right)^{-1/3}\,.
   \label{distance_gap_size}
    \end{align}
        We expect the secondary component to be orbiting with a Keplerian velocity on a circular trajectory. For the purpose of creating an empty ring between the inner and outer sub-disc, i.e. a trailing gap behind the moving secondary component (see Figure \ref{fig:disc_gap_ADAF.pdf}), the viscous time-scale of the accretion disc has to be larger than the orbiting period of the secondary component
        \begin{equation}
            \frac{t_{\rm{vis}}}{T_{\rm{orbit}}} = \frac{R^2}{H^2}\frac{1}{2\alpha\pi}.
        \end{equation}
        The disc thickness plays a role in the perturber's ability to create a stable trail when orbiting around the primary body. Hence, we assume the accretion disc scale-height to radius ratio  ${H}/{R} \ll 1$ and  $\alpha \leq 1$, corresponding to the standard thin Shakura-Sunyaev accretion, and the ratio  $t_{\rm{vis}}/T_{\rm{orbit}}$ is then much greater than unity.

We have also taken into account the possible star-disc interactions with a perturber as a massive star \citep{Davies_2020}. However, the influence of such perturbers on the simulated spectral properties is expected to be negligible.
The cavity due to the presence of a corotating or a counter-rotating star can be formed via the gravitational or the magnetohydrodynamical interaction. The gravitational effect can be estimated with eq.~(\ref{distance_gap_size}): the gap width in the system with a primary SMBH of mass $10^9 M_{\odot}$ and a perturbing star of $100M_{\odot}$ comes out to be 
\begin{equation}
    \frac{\Delta R_{\star}}{R_{\rm g}}\approx 0.64 \left(\frac{d}{100R_{\rm g}}\right)\left(\frac{M_{\star}}{100\,M_{\odot}}\right)^{1/3}\left(\frac{M_1}{10^9\,M_{\odot}}\right)^{-1/3}\,,
\end{equation}
hence generally less than $1\,R_{\rm g}$ for main-sequence stars.

The hydrodynamical interaction of embedded wind-blowing stars with the surrounding accretion disc is expected to form cavities that are nearly spherical for coorbiting stars. For the counter-orbiting stars, cavities are comet-shaped, i.e. elongated along the relative motion. The characteristic radius of such a cavity is given by the stagnation radius \citep{1996ApJ...459L..31W,2016MNRAS.455.1257Z,2021ApJ...917...43S}, 
\begin{equation}
    R_{\rm stag,\star}=\left[\frac{\dot{m}_{\rm w}v_{\rm w}}{4\pi \rho_{\rm disc}(v_{\rm rel}^2+c_{\rm s}^2)}
    \right]^{1/2}\,,
    \label{eq_stag_wind}
\end{equation}
where $\dot{m}_{\rm w}$ and $v_{\rm w}$ are the wind mass-loss rate and the wind terminal velocity, respectively. The radial mass density profile of the standard thin disc $\rho_{\rm disc}$ is given by the Shakura-Sunyaev solution \citep{Frank}. The relative velocity of the star orbiting at distance $r_{\star}$ from the SMBH is nearly zero for coorbiting stars, $v_{\rm rel}\approx 0$, while it is $v_{\rm rel}\approx 2\sqrt{GM_1/r_{\star}}$ for counter-orbiting stars. The sound-speed profile in the disc $c_{\rm s}(T_{\rm disc})$ is given by the standard Shakura-Sunyaev solution as well \citep{Frank}. { In Fig.~\ref{fig_stag_radius_star}}, we show the stagnation radii (expressed in gravitational radii) for coorbiting and counter-orbiting wind-blowing stars in the left and the right panels, respectively.
For the coorbiting case, the stagnation radii are at most $\sim 10^{-2}R_{\rm g}$ for a rather extreme mass-loss rate of $\sim 10^{-4}-10^{-3}\,{\rm M_{\odot}\,yr^{-1}}$. For counter-orbiting stars, the stagnation radii are approximately three orders of magnitude smaller for the same parameter range.

Young energetic and strongly magnetized neutron stars (or pulsars if the magnetic axis sweeps across the line of sight) are sources of the relativistic wind of accelerated particles that transfer momentum to the surrounding accretion disc. Again, this leads to the formation of the cavity for pulsars that are either coorbiting or counter-orbiting with the surrounding accretion disc. The contact discontinuity forms between the shocked pulsar wind and the shocked disc material. The characteristic radius is given by the balance between the pulsar wind pressure on one side and the ram pressure as well as the disc gas thermal pressure on the other side \citep{2015AcPol..55..203Z,2021ApJ...917...43S}, 
\begin{equation}
    R_{\rm stag,psr}=\left[\frac{\dot{E}_{\rm sd}}{4\pi \rho_{\rm disc}c(v_{\rm rel
    }^2+c_{\rm s}^2)} \right]^{1/2}\,,
    \label{eq_psr_wind}
\end{equation}
where $\dot{E}_{\rm sd}$ is the pulsar spin-down energy that we consider in the range of $10^{30}-10^{39}\,{\rm erg\,s^{-1}}$. { In Fig.~\ref{fig_stag_radius_psr}}, we show the dependence of $R_{\rm stag,psr}$ on the pulsar distance from the SMBH and on the spin-down energy for the coorbiting and the counter-orbiting case in the left and right panels, respectively. 
For the thin disc environment, the largest cavities are expected to have $R_{\rm stag, psr}\lesssim 10^{-3}R_{\rm g}$ for the coorbiting case, while the stagnation radius is three orders of magnitude smaller for the counter-orbiting case. The following analysis shows that gaps with sizes of $2R_{\rm stag}\ll 1\,R_{\rm g}$ have a negligible effect on the SED of the perturbed thin disc. However, the case of a hot flow is different and cavity and bow-shock sizes will be larger depending on the accretion rate. For inclined orbits, this can lead to quasiperiodic behaviour of both the inflow and the outflow \textcolor{black}{ rates} \citep{2021ApJ...917...43S}.

\textcolor{black}{ Considerations above, as well as the rich literature addressing the issue of the gap opening in the disc \citep[e.g.][]{1986ApJ...309..846L,2020ApJ...904..121M,1993MNRAS.264..388K,1996ApJ...460..832T,1996PhRvD..53.2901C,1997Icar..126..261W,1999A&A...352..452S,1999A&A...341..385C,2000ApJ...536..663N,karas,2021MNRAS.501.3540D,2021MNRAS.505.1324P} shows that there is no way to evaluate the fraction of the material lost when crossing the gap either to an outflow, or to a secondary body. This would have to be a new independent phenomenological parameter, and our pilot study does not justify the introduction of such a parameter. This could be done when the specific high-precision data set is considered.}

\subsection{Intrinsic reddening}

\textcolor{black}{ We also aim to test if the possible intrinsic reddening affects the recovery of the disc and gaps parameters. Intrinsic reddening changes the spectral slope but the modification is potentially different from the modification introduced by the inner gap. The intrinsic reddening in extragalactic sources is in general quite different from that described by the standard extinction curve proper for the interstellar medium of the Milky Way studied by \citet{seaton1979} and \citet{cardelli1989}. For example, even the extinction in Small and Large Magellanic Clouds differs from the Milky Way dust properties \citep[e.g.][]{gordon2003}. In general, other active or star-forming galaxies show no strong 2175 \AA~ graphite feature \citep[e.g.][]{calzetti1994}. A number of extinction curves for AGN was proposed \citep[e.g.][]{czerny_li2004,gaskell2004,zafar2015}, including a simple use of the extinction curve from the Small Magellanic Cloud advocated by \citet{hopkins2004}. The question which of the available extinction curves is preferred has not yet been fully resolved \citep{fawcet2022}.}

\textcolor{black}{ For the current study we use the simple extinction curve from \citet{czerny_li2004} in its analytical form}
\begin{equation}
\log \nu L_{\nu}^{\rm reddened} = \log \nu L_{\nu} - 0.4 A_{\lambda},
\end{equation}
\textcolor{black}{ where $\lambda$ refers to the current wavelength of the photon corresponding to $\nu$, and $L_{\nu}$ vs.\ $L_{\nu}^{\rm reddened}$ refer to the intrinsic vs.\ dust-attenuated source luminosity. We apply the reddening to the spectrum in the rest frame as}
\begin{equation}
\frac{A_{\lambda}}{E(B-V)}\left\{
\begin{array}{lll}
 & \!\!\!\!\!\! = -1.36 + 13 \log \frac{1}{\lambda} & \!\!\!\!\!\!{\rm for}\; 1.5 < \lambda^{-1} <8.5\; [\mu m^{-1}],  \\[3pt]
 & \!\!\!\!\!\! = \; 0   & {\rm for}\; \lambda^{-1} < 1.5\; [\mu m^{-1}],\\[3pt]
 & \!\!\!\!\!\! = -1.36 + 13 \log(8.5)  & {\rm for}\; \lambda^{-1} > 8.5\; [\mu m^{-1}].  
\end{array} \right. 
\label{eq:extinction}
\end{equation}
\textcolor{black}{ The constant $E(B-V)$ is the model parameter that measures the amount of dust along the line of sight.}

\section{Results}
\label{sec_results}

In Subsection~\ref{sect:outer_gap_justification}, we present three basic perturbations of the standard thin disc: the inner cavity due to the transition to the ADAF (model A), the gap created by an orbiting perturber - a secondary massive black hole (model B), and their combination, which naturally results in a two-gap setup (model C). A standard thin disc still serves as a useful comparison model. To make a connection to real observations, in Subsection~\ref{sect_models} we generate synthetic data according to model A and B scenarios and assess how precisely the adopted model parameters can be recovered for the fiducial data accuracy of \textcolor{black}{ 0.125 mag and 0.025 mag, corresponding to the accuracy of  5\% and 1\% in the measured flux, respectively.} 

\subsection{Adopted parameters and justification of the assumptions}
\label{sect:outer_gap_justification}

The free parameters of the multi-gap model are: inclination $i$, mass of the primary (central) black hole $M_{1}$, the accretion rate of the primary black hole in Eddington units $\dot{M_{\bullet}} = \dot{m} M_{\rm{Edd}}$, \textcolor{black}{ the inner radius of the disc set by hot inner ADAF flow, $R_{ADAF}$,} the mass of the secondary black hole $M_{2}$ , and the  distance between the primary (central) and the secondary black hole $d$. The latter two parameters $M_{2}$, and $d$ apply only in case of the presence of the secondary component in the system. We employ the color correction factor $f_{\rm{col}}$ values between 1 and 2 ($f_{\rm{col}} = 1.6$) to our local black-body emission from the disc. 

\begin{figure}
    \centering
    \includegraphics[width=\columnwidth]{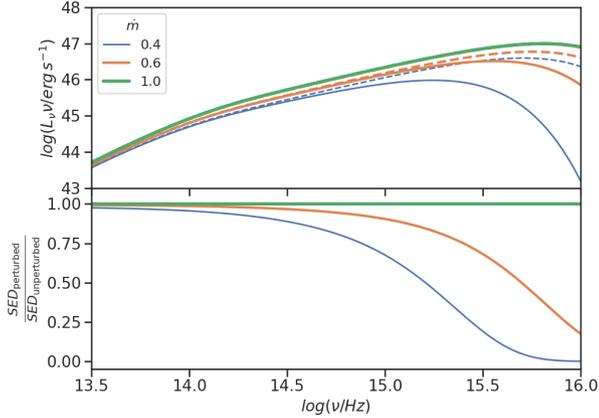}
    \caption{Simulated SED from an accretion disc with a inner gap (top panel; the model A), viewed at the inclination of 35$\deg$ \textcolor{black}{ under the assumption of the strong ADAF principle}. \textcolor{black}{ The mass of the primary is set to $10^9M_{\odot}$.} Different colors correspond to the dimensionless mass accretion rate (scaled with respect to the  Eddington rate). Solid and dashed lines mark the perturbed and the unperturbed SED, respectively. In order to assess the effect of the gap, we also show the ratio of the perturbed with respect to the unperturbed model (bottom panel).}
    \label{fig:adaf.png}
    
\end{figure}

 \textcolor{black}{ In this work, we neglect the radiative contribution of the ADAF component to the SED. Using eq. (\ref{uv_xray_rel}) we can estimate the ADAF contribution to the total spectrum not exceeding 1\% in case $\dot{m}$ set to 0.4. Consequently, we are allowed to ignore the ADAF contribution to the optical/UV part of the SED spectra (see Figure \ref{fig:adaf.png}).}

\textcolor{black}{ The number of the free parameters in the model is large, and we cannot constrain all of them independently and uniquely. In the case of only the inner gap present, the accretion disc spectrum like in a standard Shakura-Sunyaev disc is actually a two-parameter model, since the shape of the spectrum scales with a combination of parameters. All spectra are identical in the log-log plane apart from vertical and horizontal translation, and the presence of the truncation radius does not change this property. Therefore, only two out of four parameters  $M_{1}$, $\dot{M_{\bullet}}$, $i$, $R_{ADAF}$, can be constrained. The value of the black hole mass can be derived independently, from the width of the spectral lines \citep[e.g.][]{Vestergaard2006,mejia2018}.  We need one more constraint, and we consider two alternatives.}

\textcolor{black}{ As the first option, in order } to describe the inner gap in the disc, we follow the so-called \textit{strong ADAF principle}: whenever the ADAF solution is allowed at a certain radius $R$, i.e. when the local accretion rate drops below the critical value, the flow proceeds in this regime \citep{1995ApJ...438L..37A,1996PASJ...48...77H,1998PASJ...50..559K}. For sources that accrete close to the Eddington limit, a standard thin disc extends essentially all the way to the ISCO. For lower accretion rates, an inner region turns into the hot diluted ADAF. The separation radius between the ADAF and the thin disc is given approximately by \citep{1995ApJ...438L..37A,1996PASJ...48...77H,1998PASJ...50..559K},
    \begin{equation}
        R_{\rm ADAF}=4\alpha_{0.1}^4\dot{m}^{-2}R_{\rm g}.
        \label{r_adaf_eq}
    \end{equation}
    where $\alpha_{0.1}$ is the viscous parameter expressed in 0.1 and $\dot{m}$ is the dimensionless accretion rate expressed in Eddington units, $\dot{m}=\dot{M_{\bullet}}/\dot{M}_{\rm Edd}$. The Eddington accretion rate is given by
    \begin{equation}
        \dot{M}_{\rm Edd}=\frac{48\pi GM_{\bullet}m_{\rm p}}{\sigma_{\rm T}\,c}\,,
    \end{equation}
    where we assumed the radiative efficiency of $\eta=1/12$.
    For instance, for the accretion rate of $\dot{m}=0.4$, $R_{\rm ADAF}\sim 25 R_{\rm g}$. The width of the gap in the standard thin disc is then given by $R_{\rm ADAF}-R_{\rm ISCO}$, that is $\sim 19$ $R_{\rm{g}}$ for a non-rotating black hole.

\textcolor{black}{ Such an assumption leads to a very strong dependence on}
    the dimensionless accretion rate (see Figure \ref{fig:adaf.png}) which leads to the change of the values of the inner radius of the accretion disc via eq. (\ref{r_adaf_eq}). 
    
    The residuals show that the more the relative accretion rate decreases, the more the perturbed SED deviates from the unperturbed one. This effect reflects 
    the fact that the accretion disc, being the source of the thermal component, is getting smaller as the inner hot flow not contributing to the UV bands expands. Therefore the SED simulation of the thermal component shows a sharp flux decrease in the high frequency range.

  \begin{figure}
    \centering
    \includegraphics[width=\columnwidth]{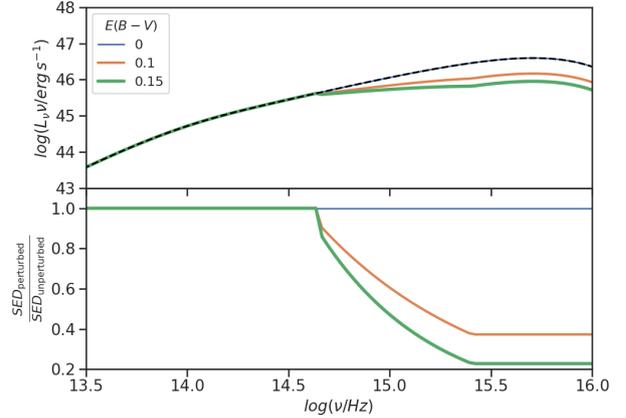}
    \caption{\textcolor{black}{ Simulated SED from an accretion disc with dust component present (top panel), viewed at the inclination of 35$\deg$ . \textcolor{black}{ The mass of the primary is set to $10^9M_{\odot}$.}Different colors correspond to the extinction $E(B-V)$ (see eq. (\ref{eq:extinction})). Solid and dashed lines mark the perturbed and the unperturbed SED, respectively. In order to assess the effect of the dust component, we also show the ratio of the perturbed with respect to the unperturbed model (bottom panel).}}
    \label{fig:dust.png}
\end{figure}
    
\textcolor{black}{ As the second option, we consider the viewing angle $i$ to be known. In fact, there are methods to determine $i$ from the data (e.g. from the broad band data fitting, including X-rays and IR, e.g. \citealt{prince2022,ramos2022}, from the dynamics of the narrow line region, e.g. \citealt{fischer2013}, polarimetry, e.g. \citealt{marin2014}, and eventually from the [OIII] emission, although this seems problematic, e.g. \citealt{diamond2009}), and statistically, this viewing angle centers at about $\sim 35$ deg due to the presence of the dusty torus which shields the view of the nucleus for highly inclined objects \citep[e.g.][]{lawrence2010,prince2022}. When the effects of  varying the accretion rate and the inner radius are decoupled, the changes with each of the parameters are slower than in Figure~\ref{fig:adaf.png}, but qualitatively similar. }

\textcolor{black}{ When we consider the dust reddening, we see that the turn-off at high frequencies caused by the extinction and the turn-off caused by $R_{ADAF}$ have different shapes. The turn-off is more shallow, and in addition a small discontinuity in the spectrum slope appears due to the formula used at the shortest wavelengths (see Equation~\ref{eq:extinction}). This causes a turn-up in the far-UV (see Figure \ref{fig:dust.png}). On one hand, perhaps the transition to saturation could be done more smoothly. On the other hand, such a turn-up  is actually observed in some sources \citep[e.g.][]{lawrence2012,marculewicz2022}, or needed to represent well the unobserved ionizing continuum \citep[][]{binette2022}.}

\begin{figure}
        \centering
        \includegraphics[width=\columnwidth]{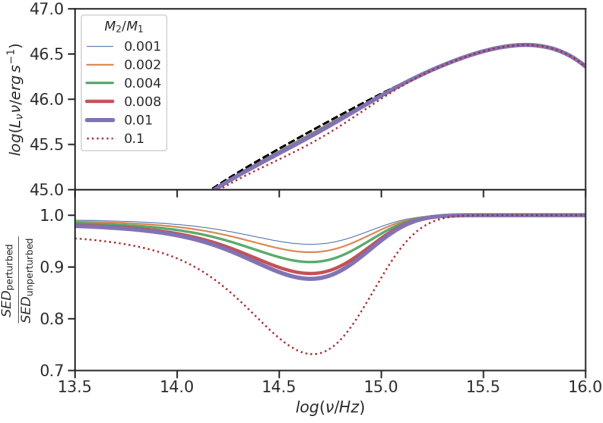}
        \caption{Simulated SED from an accretion disc with an outer gap (top panel; the model B), viewed at the inclination of 35$\deg$. \textcolor{black}{ The mass of the primary is set to $10^9M_{\odot}$.} Different colors correspond to the mass of the secondary component (scaled with respect to the mass of the primary body and located at a fixed distance $d$ set to 400$R_{\rm{g}}$) which translates into the respective gap width given via eq. (\ref{distance_gap_size}). Solid and dashed lines mark the perturbed and the unperturbed SED, respectively. In order to assess the effect of the gap
        , we also show the ratio of the perturbed with respect to the unperturbed model (bottom panel). We involve the SED simulations for the mass of the secondary component \textcolor{black}{ from 0.001$M_{1}$
        } up to 0.1$M_{1}$ (indicated by a dotted line) only for the illustrative purposes as we do not consider such a scenario further due to the disagreement with our assumptions about the studied system (see subsect.~ \ref{sect:outer_gap_justification}). }
        \label{fig:sec_BH_revised_masses}
    \end{figure}

\textcolor{black}{ In the case of the outer gap, in order to test the interesting range of the secondary black hole mass we performed the computations for a range of masses, all located at the same distance from the primary black hole. We consider the mass ratio from 0.1 down to 0.001. The result is shown in Figure~\ref{fig:sec_BH_revised_masses}.}

    \textcolor{black}{ Later on we} do not consider the possibility of a secondary component having the mass $M_{2} \approx M_{1}$ and $M_{2} \approx 10^{-1} M_{1}$ as this would result in both the primary and the secondary components orbiting their centre of mass, with the secondary at the distance of $r_2=r/2$ and $r_2\sim 0.9r$, where $r$ is the primary-secondary distance, respectively \citep[e.g.][]{Gold_2014}. For the case of nearly equal-mass binaries, the MHD numerical simulations show that each of the black holes would be surrounded by an additional small mini-disc \citep[e.g.][]{minidiscs_1, minidiscs_2}, which would represent very different geometrical set-up compared to the one considered here.
    
    \textcolor{black}{ Too small mass of the }secondary component with $M_{2} \approx 10^{-3} M_{1}$ would, on the other hand, account for rather small  or non-existent gaps; see eq.~(\ref{distance_gap_size}). This would translate into a negligible decrease of the monochromatic luminosity up to $\sim 5\%$ (see Figure \ref{fig:sec_BH_revised_masses}).
    \textcolor{black}{ We set the mass of the primary to $M_{1}$=$10^{9}M_{\odot}$ for definiteness of the example. Finally, in the context of the model B and C requiring the presence of a secondary body, we consider $M_{2}=10^7M_{\odot}$, yielding the mass ratio $0.01$, unless stated otherwise (see Figure \ref{fig:sec_BH_revised_masses}).}

\begin{figure}
    \centering
    \includegraphics[width=\columnwidth]{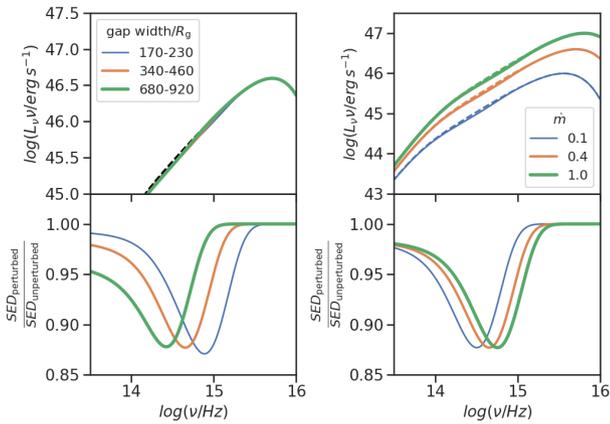}
    \caption{Left panel: SED simulated corresponding to the model model B with varying gap width with the secondary component located in the middle of the gap. Other parameters: inclination angle has been set to 35$\deg$; the dimensionless accretion rate $\dot{m}$ is 0.4 with respect to the Eddington value. \textcolor{black}{ The mass of the primary is set to $10^9M_{\odot}$, mass of the secondary set to $10^7M_{\odot}$.} Black dashed line marks the unperturbed SED. Maximum drop of the SED ratio reaches about 12$\%$ for the largest width of the gap shown (left panel). Right panel: Simulated SED corresponding to model B with parameters: inclination set to 35$\deg$,  distance between the primary and secondary body set to $400 R_{\rm{g}}$ with the respective gap width set to 120$R_{\rm{g}}$. \textcolor{black}{ The mass of the primary is set to $10^9M_{\odot}$, mass of the secondary set to $10^7M_{\odot}$.} The set-up of the graphs is analogous to the previous figure. Colored dashed lines mark an unperturbed spectral energy density (right panel).}
    \label{sec_BH_revised}
\end{figure}

\begin{figure}
    \centering
    \includegraphics[width=\columnwidth]{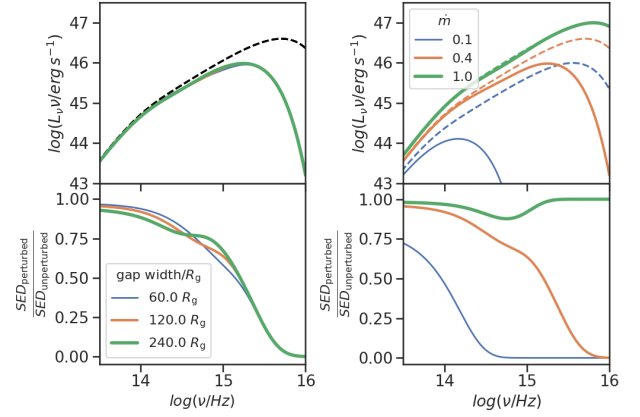}
    \caption{Simulated SED corresponding to the model C \textcolor{black}{ under the assumption of strong ADAF principle}. 
    Left panel: the case of a fixed accretion rate $\dot{m}$ is 0.4 and the inclination angle $35\deg$. \textcolor{black}{ The mass of the primary is set to $10^9M_{\odot}$, mass of the secondary set to $10^7M_{\odot}$. The gap width varies with the distance of the secondary component located in the middle of the gap, corresponding to 200, 400 and 800$R_{\rm{g}}$, respectively.} Black dashed line marks the unperturbed SED for reference. The gradual development of a bump occurs as the gap widens. Right panel: inclination has been set to 35$\deg$, distance between the primary and secondary body set to $400 R_{\rm{g}}$ with the respective gap width set to $120 R_{\rm{g}}$. \textcolor{black}{ The mass of the primary is set to $10^9M_{\odot}$, mass of the secondary set to $10^7M_{\odot}$.} Colored dashed lines indicate the corresponding unperturbed SED profiles.}
    \label{adaf+sec_BH_revised}
\end{figure}

    In case of model B, 
    we turn off the inner gap option, assuming the presence of the secondary component only, and we vary the size of the gap created within the accretion disc.
    The size of the gap created by the secondary black hole shows a linear dependence on its distance from the primary one; see eq. (\ref{distance_gap_size}). That leads to the conclusion that bigger gaps located further away will have an effect on the continuum profile mainly in the lower-frequency range, and vice versa, see Figure~\ref{sec_BH_revised} (left panel). However, changing the value of the relative accretion rate shows that the gap presence will eventually have an effect on the continuum profile in the high-frequency interval as well (Fig.~\ref{sec_BH_revised}, right panel). That is a general result of the SED peak being shifted to higher energies as its accretion rate approaches the Eddington limit. We also find that the color correction value does not influence the effect of the gap on the SED. The only qualitative change in the spectra is the shift to the higher-frequency range.  
    
    For model C, we first fix the size of ADAF region and keep varying the size of the gap formed by the secondary black hole. We notice that the gap width must be sufficiently big in order to outweigh the changes to the SED caused by the presence of ADAF region, see Figure \ref{adaf+sec_BH_revised} (left panel). The effect of the secondary component is clearly observable for the accretion disc with the relative accretion rate close to unity, see Figure \ref{adaf+sec_BH_revised} (right panel), i.e. highly accreting sources. Once the relative accretion rate approaches $\sim 0.1$, the ADAF region \textcolor{black}{ under the assumption of the strong ADAF principle}  becomes so extended that its effect merges with the effect of the gap, which makes the gap influence on the continuum indistinguishable.
    
    We are limited to $R_{\rm{ISCO}} = 6 R_{\rm{g}}$, i.e. for the highly accreting sources, such as sources with $\dot{m} = 1.0$, the ADAF region is then cut off.  
    One has to realize that $R_{\rm{inner}}$ can vary as well -- especially in model A involving the ADAF scenario, for which the position of the inner accretion-disc radius is shifted. Such a behaviour would be independent of the number of gaps and would result in a different temperature profile in eq. (\ref{eq_temperature}) and the lower limit of the integral in eq. (\ref{eq_luminosity}) as well as in eq. (\ref{gap_luminostiy}).
    
\subsection{Models compared to synthetic data} 
\label{sect_models}

\textcolor{black}{ Now we test if the expected gaps in the cold standard accretion disc can be detected in the actual data. Massive photometric data are available or will be available for numerous sources, as a result of the all sky survey but the issue is whether the departure due to the gap can really be detected in such data. Therefore, we select all sky survey instruments, create artificial photometric data out of our continuous models, assign the photometric error and later, by data fitting, check if the model parameters can be recovered.}

\textcolor{black}{ We use the photometric system of the GALEX \citep{morrissey2007}, SWIFT/UVOT \citep{poole2008}, Vera Rubin Observatory which will provide the Legacy Survey of Space and Time (LSST) \citep{LSST_SB2009,huber2021}, and we supplement them with the two colors W1 and W2 from WISE, using just the effective wavelegths\footnote{\url{https://www.astro.ucla.edu/~wright/WISE/passbands.html}}; see also the instrument webpage for more detailed information\footnote{\url{https://wise2.ipac.caltech.edu/docs/release/allsky/expsup/sec4\_4h.html\#WISEZMA}}. We did not include W3 and W4 since for low redshift sources these bands are dominated by the host galaxy. They might be useful for higher redshift sources but then assigning the host contamination would complicate the issue. The effective wavelengths are given in Table~\ref{tab:instruments}. We also include there the width of the filter although we do not fold the model spectrum with the filter profile in the current study.}

\textcolor{black}{ To justify the combination of points from multiple instruments (see Table \ref{tab:instruments}), the studied source should not vary substantially within the monitored timescale, i.e. higher-accreting sources are in this sense more suitable as they are expected to have a lower fractional excess variability in comparison with low-accreting systems \citep[see e.g.][]{2021ApJ...912...10Z}. A set of joined follow-up observations, timed corresponding to the characteristic UV/optical timescale $\tau_{\rm char}$, could potentially help to counter the source  variability \citep{2001ApJ...555..775C}. For most AGN, the campaign should be within $\tau_{\rm char}\sim 30-40$ days so that the flux densities can be considered quasi-simultaneous, i.e. corresponding to the same accretion state. In addition, heavier SMBHs of $\sim 10^9\,M_{\odot}$ assumed in this work are more suitable because of proportionally longer variability timescales, which are between about one day and about two months considering different timescales at the emission distance of $20\,R_{\rm g}$ \citep{2001ApJ...555..775C},
\begin{align}
    t_{\rm lc} &=1.10 \left(\frac{M_{\bullet}}{10^9\,M_{\odot}} \right)\left(\frac{R}{20\,R_{\rm g}} \right)\,\text{days}\,\notag\\
    t_{\rm acc} &\geq 5.06 \left(\frac{M_{\bullet}}{10^9\,M_{\odot}} \right)\left(\frac{R}{20\,R_{\rm g}} \right)^{3/2}\,\text{days}\,\notag\\
    t_{\rm orb} &= 33.0 \left(\frac{M_{\bullet}}{10^9\,M_{\odot}} \right)\left(\frac{R}{20\,R_{\rm g}}\right)^{3/2}\,\text{days}\,\notag\\
    t_{\rm th} &= 53.0 \left(\frac{\alpha}{0.1} \right)^{-1} \left(\frac{M_{\bullet}}{10^9\,M_{\odot}} \right)\left(\frac{R}{20\,R_{\rm g}} \right)^{3/2}\,\text{days}\,,
\end{align}
where $t_{\rm lc}$, $t_{\rm acc}$, $t_{\rm orb}$, $t_{\rm th}$ denote light-crossing, accretion (free-fall), orbital, and thermal timescales.
There is also a possibility of including data from the future detectors such as UVEX\footnote{\url{https://www.uvex.caltech.edu/page/about}},  which has been proposed to cover a similar FUV band like GALEX with approximately two orders of magnitude better sensitivity, or the data from current detectors such as AstroSat/UVIT  which could provide further improvement in S/N \citep[e.g.][]{2021MNRAS.504.4015D,2023arXiv230311882K}. In addition, planned small versatile UV satellites such as ULTRASAT \citep{2022SPIE12181E..05B} or QUVIK \citep{2022arXiv220705485W} will be flexible enough to join in the campaigns to provide quasi-simultaneous flux densities in the near-UV and far-UV bands. In summary, there seem to be very good prospects for improving the above-described analysis in the future.}

\textcolor{black}{
 In our simulations, we assume the sources to be of high luminosity and we do not take into account additional contribution of Balmer continuum as well as the contribution of strong BLR lines. These spectral components should be considered in the improved theoretical model of an accretion disk with gaps or alternatively, these components have to be considered in the spectral data analysis to extract continuum flux densities for specific sources.}

\textcolor{black}{ The mock data points are then corrected to include the photometric error. We do that assuming the error level and applying the Gaussian random distribution around the adopted mean value. In the current study, we assume the same level of the photometric error in all bands. Such a spectrum represents the rest-frame model. We also consider mock sources located at different redshifts, in which case the continuum model is redshifted first and then the photometric data are generated in the observed frame.}

\textcolor{black}{ The mock data is then fitted using the model, with the aim to check if the adopted parameters are recovered for the considered setup. We do this using $\chi^2$ fitting unless stated otherwise. }

\textcolor{black}{ 
As discussed in Section~\ref{sect:outer_gap_justification} in the case of the model A data we cannot fit all four parameters at the same time. Our fit examples have been made for a fixed black hole mass and viewing angle, varying $\dot m$ and $R_{\rm ADAF}$. 
The results are shown in Table \ref{synth_fit_A}.}

\textcolor{black}{ In case of fitting the model B data we proceed with the fixed parameters -- $M_{1}$, $i$, $R_{\rm{ in}}$ and $R_{\rm{outer}}$. The only free parameters of the model are the $\dot{m}$,   $R_{\rm{gap\, in}}$ and $R_{\rm{gap\, out}}$. To justify our choice of fixed and free parameters we argue that $M_{1}$ and $i$ can be inferred by the study of the changes in the UV/optical emission line and the polarized continuum radiation, respectively.  }

\begin{table}
\centering
\begin{tabular}{|c|c|c|c|}

\hline
 Instrument & Filter & \makecell{Central Wavelength\\($\textup{\AA}$)}  & \makecell{FWHM/bandpass\\($\textup{\AA}$)}    \\ \hline
 \multirow{ 2}{*}{GALEX} & NUV & 2315.7 & 1771 -- 2831   \\ 
 & FUV & 1538.6 & 1344 -- 1786  \\ \hline 
 \multirow{ 6}{*}{LSST} &  u & 3671 &  3200 - 4000  \\ 
 & g & 4827 &  4000 -  5520  \\ 
  & r & 6223 &  5520 -  6910  \\ 
 & i & 7546 &   6910 -  8180  \\ 
  & z & 8691 &  8180 -  9220  \\ 
 & y & 9712 &   9500 -  10800 \\ \hline 
 \multirow{ 6}{*}{SWIFT} & v & 5468 & 769   \\ 
 & b & 4392 & 975   \\ 
   & u & 3465 & 785   \\ 
 & uvw1 & 2600 & 693  \\ 
   & uvm2 & 2246 & 498   \\ 
 & uvw2 & 1928 & 657   \\ \hline 
 \multirow{ 2}{*}{WISE} & w1 & 33680 & 28000 - 38000\\
 & w2 & 46180 &  41000 - 51000\\
\hline

\end{tabular}
\caption{ Photometry table illustrating the central wavelength alongside the FWHM (in some cases) for the given instrument's channel.}
\label{tab:instruments}
\end{table}

\par
\textcolor{black}{ We present the best-fit results for the simulation models A and B with errors corresponding  5\% in the measured flux. In Table \ref{synth_fit_A}, we show that larger redshift helps to uncover the free parameters of the model, i.e. the bigger the redshift, the better the reduced chi-square value $\chi_{\nu}^{2}$ as well as the fitted values are corresponding to the assumed ones. That is expected as the SED differences between the perturbed and unperturbed model start to be visible for high frequencies. }

\textcolor{black}{ Consequently, we add the dust component to model A and show the fitting results for our mock data (see Table \ref{synth_fit_dust}). Given the errors up to 5\% in measured flux, we were able to recover the free parameters, including the extinction $E(B-V)$. Specifically, the bigger the redshift, the better fit results we achieved.}

\textcolor{black}{ In case of model B, we fit each of our mock data set with either $\dot{m}$, $R_{\rm{gap\,in}}$, $R_{\rm{gap\,out}}$ as free parameters (secondary gap scenario) or with the $R_{\rm{gap\,in}}$ and $R_{\rm{gap\,out}}$ fixed both to the same value (no gap scenario), i.e. first row in each cell corresponds to the fitting run using perturbed model and the second row  corresponds to the fitting run using unperturbed model (see Table \ref{synth_fit_B}). }

\textcolor{black}{ We apply both the Akaike \citep[$AIC$,][]{akaike1973information} and Bayesian Information Criterion \citep[$BIC$,][]{1978AnSta...6..461S}, as a standard statistical test method for selection among the finite set of models, in our case the unperturbed vs. perturbed ones. }
\textcolor{black}{
The $\chi_{\nu}^{2}$ values for the model B scenario in Table \ref{synth_fit_B} show the fitting results with or without the gap do not lead to significant changes. The $\Delta AIC$ and $\Delta BIC$ for each simulation are ambiguous and they show that the perturbed model can also be mildly favored. We try to infer the mass of secondary as well and its distance using the fit results. The values we recover indicate that the mass of secondary is even consistent with zero. The gap creates only a small depression, up to $\approx$ 12\% at just one frequency (e.g. see Figure \ref{sec_BH_revised}, and much less at other frequencies, so 5\% error potentially hides the difference. 
}

\textcolor{black}{ Therefore, we fit the same mock data but now with 1\% uncertainty (see Table \ref{synth_fit_B2}). Comparing the results for $\chi_{\nu}^{2}$, $\Delta AIC$, and $\Delta BIC$ characteristics, we verify that the perturbed model is favoured for all mock data sets and we are able to recover the initial parameters of mass and distance of the secondary with good precision. As expected, the fit results become worse when a higher-redshift object is considered. That is expected since the gradual shift to a higher redshift causes SED changes to become less covered, since in the majority of the considered cases, they are present in the low-frequency range.}

\begin{figure}
    \centering
    \includegraphics[width=\columnwidth]{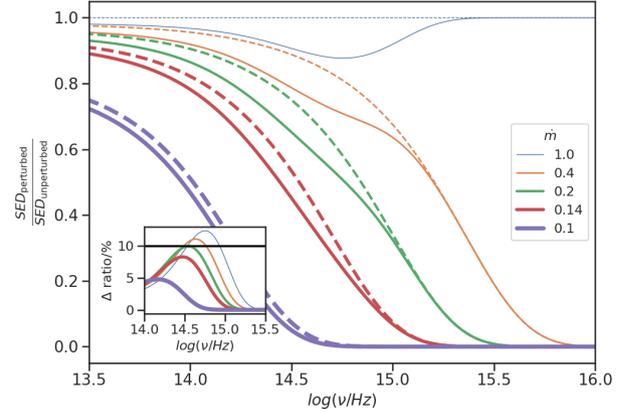}
    \caption{Simulated SED model ratios for model A (dashed lines) and model C (solid lines)\textcolor{black}{, both under the assumption
of strong ADAF principle,} with the following fixed parameters: the inclination angle set to 35 $\deg$, distance between the primary and secondary body set to $400 R_{\rm{g}}$ with the respective gap width set to 120$R_{\rm{g}}$. \textcolor{black}{ The mass of the primary is set to $10^9M_{\odot}$, mass of the secondary set to $10^7M_{\odot}$.} Lines are color-coded by $\dot{m}$. The main plot shows the ratio between the cases with and without the gap in the accretion disc. The inset shows the corresponding difference between the model ratios (the black solid line marks the 10\,\% threshold).}
    \label{model_C_comparison}
\end{figure}

Finally, the model C spectra are essentially a result of a combination of both models A and B\textcolor{black}{ , under the assumption
of strong ADAF principle.} However, as the size of ADAF region is different, the effect of the gap caused by the secondary body gets negligible (see the right panel in Figure \ref{adaf+sec_BH_revised}). For the sources with a low accretion rate, the size of the ADAF could potentially reach the gap itself or overlap it. A basic comparison of model A and C ratios between the perturbed and the unperturbed SEDs (see Figure \ref{model_C_comparison}) provides us with the constraint on the lower limit of the accretion rate, for which the inner ADAF does not prevent the gap detection due to the secondary black hole at the fixed distance of $d$ set to 400 $R_{\rm{g}}$, which corresponds to the size of the gap 340--460$R_{\rm{g}}$. For the accretion rate $\dot{m}>0.2$, we observe both the ADAF as well as the secondary black hole effect, assuming an arbitrary $10\%$ threshold for the difference between the two models in the narrow frequency range $\approx 10^{14.5}$--$10^{15.0}$Hz. 

\begin{table*}
\centering
\begin{tabular}{ c|c|c|c|c }

\hline
 \hline 
  \multicolumn{2}{c|}{assumed values}& \multicolumn{3}{c}{fit values} \\ \hline
    \makecell{$\dot{m}$\\ } & \makecell{$R_{\rm{ADAF}}$\\ ($R_{\rm{g}}$)} &  {\makecell{$\dot{m}$\\ }} & \makecell{$R_{\rm{ADAF}}$\\ ($R_{\rm{g}}$)} & $\chi^2_{\nu}$   \\
 \hline

\multicolumn{5}{c}{z=0} \\ \hline
  0.4 & 11 & 0.37 $\pm$ 0.03 & 8 $\pm$ 3 & 1.1  \\ 
  0.4 & 25 & 0.39 $\pm$ 0.04 & 28 $\pm$ 4 & 1.8   \\ \hline

\multicolumn{5}{c}{z=0.5} \\ \hline
  0.4 & 11 & 0.40 $\pm$ 0.04 & 10 $\pm$ 2 & 1.8   \\
  0.4 & 25 & 0.41 $\pm$  0.03 & 25 $\pm$ 2 & 1.1   \\ \hline

\multicolumn{5}{c}{z=1} \\ \hline
  0.4 & 11 & 0.38 $\pm$ 0.03 & 10 $\pm$ 1 & 1.3   \\
  0.4 & 25 & 0.39 $\pm$ 0.04 & 25 $\pm$ 2 & 1.9   \\ \hline

\multicolumn{5}{c}{z=1.5} \\ \hline
  0.4 & 11 & 0.38 $\pm$ 0.03 & 10 $\pm$ 1 & 1.7    \\
  0.4 & 25 & 0.39 $\pm$ 0.03 & 25 $\pm$ 1 & 1.1   \\  \hline

\multicolumn{5}{c}{z=2} \\ \hline
  0.4 & 11 & 0.38 $\pm$ 0.02 & 11 $\pm$ 1 & 1.1   \\
  0.4 & 25 & 0.43 $\pm$ 0.03 & 26 $\pm$ 1 & 1   \\ 
 \hline
 \hline

 \multicolumn{3}{c}{}\\

\end{tabular}
\caption{
\textcolor{black}{ Fitting results of data simulated for model A with errors corresponding  5\% in the measured flux. Each row for the particular mock data shows the assumed values of the simulation and fitting results obtained for a given redshift. Additionally we fix $M_{1}$ to $10^9 M_{\odot}$, $i$ to 35 $\deg$, the gap width to 0 $R_{\rm{g}}$ and $R_{\rm{outer}}$ to 5000 $R_{\rm{g}}$.  We judge and compare the quality of the fit using the reduced chi-square test value $\chi^2_{\nu}$.
}}
\label{synth_fit_A} 

\end{table*}

\begin{table*}
\centering
\begin{tabular}{ c|c|c|c|c|c|c }

\hline
 \hline 
  \multicolumn{3}{c|}{assumed values}& \multicolumn{4}{c}{fit values} \\ \hline
  $E(B-V)$ &  \makecell{$\dot{m}$\\ } & \makecell{$R_{\rm{ADAF}}$\\ ($R_{\rm{g}}$)} & $E(B-V)$ & {\makecell{$\dot{m}$\\ }} &  \makecell{$R_{\rm{ADAF}}$ \\ ($R_{\rm{g}}$)} &  $\chi^2_{\nu}$  \\
 \hline

\multicolumn{7}{c}{z=0} \\ \hline
  0.15 & 0.4 & 11 & 0.11 $\pm$ 0.05 & 0.42 $\pm$ 0.04 & 20 $\pm$ 11 &  1.4  \\
  0.15 & 0.4 & 25 & 0.18 $\pm$ 0.04 & 0.38 $\pm$ 0.04 & 20 $\pm$ 11 & 1.3  \\ \hline

\multicolumn{7}{c}{z=0.5} \\ \hline
  0.15 & 0.4 & 11 & 0.12 $\pm$ 0.03 & 0.38 $\pm$ 0.04 & 18 $\pm$ 4 & 1.1   \\
  0.15 & 0.4 & 25 & 0.13 $\pm$ 0.03 & 0.36 $\pm$ 0.04 & 25 $\pm$ 5 & 1.7  \\ \hline

\multicolumn{7}{c}{z=1} \\ \hline
  0.15 & 0.4 & 11 & 0.15 $\pm$ 0.02 & 0.38 $\pm$ 0.04 & 10 $\pm$ 2 & 1   \\
  0.15 & 0.4 & 25 & 0.16 $\pm$ 0.03 & 0.40 $\pm$ 0.06 & 23 $\pm$ 2 & 2  \\ \hline

\multicolumn{7}{c}{z=1.5} \\ \hline
  0.15 & 0.4 & 11 & 0.15 $\pm$ 0.02 & 0.43 $\pm$ 0.05 & 12 $\pm$ 1 & 1.2   \\
  0.15 & 0.4 & 25 & 0.14 $\pm$ 0.02 & 0.34 $\pm$ 0.04 & 24 $\pm$ 1 & 1  \\ \hline

\multicolumn{7}{c}{z=2} \\ \hline
  0.15 & 0.4 & 11 & 0.12 $\pm$ 0.02 & 0.35 $\pm$ 0.05 & 12 $\pm$ 1 & 1.4   \\
  0.15 & 0.4 & 25 & 0.15 $\pm$ 0.01 & 0.44 $\pm$ 0.05 & 26 $\pm$ 1 & 1.1  \\ \hline
 \hline
 \hline
\multicolumn{3}{c}{}\\
\end{tabular}
\caption{\textcolor{black}{ Fitting results of data simulated for dust model with the errors of 5\% in the measured flux. Each row for the particular mock data shows the assumed values of the simulation and fitting results obtained for a given redshift. Additionally we fix $M_{1}$ to $10^9 M_{\odot}$, $i$ to 35 $\deg$, the gap width to 0 $R_{\rm{g}}$ and $R_{\rm{outer}}$ to 5000 $R_{\rm{g}}$.  We judge and compare the quality of the fit using the reduced chi-square test value $\chi^2_{\nu}$.}}
\label{synth_fit_dust}
\end{table*}

\begin{table*}
\centering
\begin{tabular}{ c|c|c|c|c|c|c|c|c|c|c|c|c }

\hline
 \hline 
 
    \multicolumn{3}{c|}{assumed values}& \multicolumn{8}{c|}{fit values} & \multicolumn{2}{c}{inferred values}\\ \hline
  \makecell{$\dot{m}$\\ } & \makecell{$R_{\rm{gap\,in}}$\\ ($R_{\rm{g}}$)} & \makecell{$R_{\rm{gap\,out}}$\\ ($R_{\rm{g}}$)} & {\makecell{$\dot{m}$\\ }}  & \makecell{$R_{\rm{gap\,in}}$\\ ($R_{\rm{g}}$)} & \makecell{$R_{\rm{gap\,out}}$\\ ($R_{\rm{g}}$)} & $\chi^2_{\nu}$ & \textit{AIC} & \textit{BIC} & $\Delta$\,\textit{AIC} & $\Delta$\,\textit{BIC} & \makecell{$d$\\ ($R_{\rm{g}}$)} & \makecell{$M_{2}$\\ ($10^7 M_{\odot}$)}  \\
 \hline

\multicolumn{13}{c}{z=0} \\ \hline
 0.4 & 170 & 230 & 0.41 $\pm$ 0.05 & 167 $\pm$ 74 & 228 $\pm$ 93 & 1 & 3.2 & 5.5 & -- & -- & 198 $\pm$ 59 & 2.21 $\pm$ 13.07 \\
  0.4 & 170 & 170 & 0.36 $\pm$ 0.01 & -- & -- &  1 & 0.8 & 1.5 & $-2.4$ & $-4$ & -- & --  \\ \hline
  0.4 & 340 & 460 & 0.37 $\pm$ 0.02 & 582 $\pm$ 284 & 820 $\pm$ 435 & 1 & 3.4 & 5.7 & -- & -- & 701 $\pm$ 260 & 2.94 $\pm$ 19.50 \\ 
  0.4 & 340 & 340 & 0.34 $\pm$ 0.02 & -- & -- & 1.1 & 3.1 & 3.8 & $-0.3$ & $-1.9$ & -- & -- \\ \hline
 0.4 & 680 & 920 & 0.39 $\pm$ 0.02 & 793 $\pm$ 215 & 1395 $\pm$ 416 & 1.1 & 4.1 & 6.4 & -- & -- & 1094 $\pm$ 234 & 12.50 $\pm$ 30.25  \\
  0.4 & 680 & 680 & 0.35 $\pm$ 0.02 & --& -- & 1.5 & 7.5 & 8.2 & 3.1 & 2.2 & -- & -- \\ \hline

\multicolumn{13}{c}{z=0.5} \\ \hline
 0.4 & 170 & 230 & 0.45 $\pm$ 0.05 & 114 $\pm$ 35 & 184 $\pm$ 49 & 1.1 & 4.3 & 6.6 & -- & -- & 149 $\pm$ 30 & 7.78 $\pm$ 20.62 \\
 0.4 & 170 & 170 & 0.37 $\pm$ 0.02 & -- & -- & 1.3 & 5.5 & 6.2 & 1.2 & $-0.4$ & -- & -- \\ \hline
 0.4 & 340 & 460 & 0.42 $\pm$ 0.02 & 467 $\pm$ 245 & 654 $\pm$ 403 & 1.1 & 4.9 & 7.2 & -- & -- & 561 $\pm$ 236 & 2.79 $\pm$ 21.36 \\ 
 0.4 & 340 & 340 & 0.40 $\pm$ 0.02 & -- & -- & 1.2 & 3.5 & 4.3 & $-1.4$ & $-2.9$ & -- & -- \\  \hline
 0.4 & 680 & 920 & 0.37 $\pm$ 0.02 & 994 $\pm$ 744 & 757 $\pm$ 567 & 1.1 & 4.3 & 6.6 & -- & -- & 876 $\pm$ 468 & -1.49 $\pm$ 17.78 \\
 0.4 & 680 & 680 & 0.38 $\pm$ 0.02 & -- & -- & 1 & 1.4 & 2.2 & $-2.9$ & $-4.4$ & -- & -- \\ \hline

\multicolumn{13}{c}{z=1} \\ \hline
 0.4 & 170 & 230 & 0.38 $\pm$ 0.03 & 157 $\pm$ 107 & 215 $\pm$ 149 & 1.8 & 12.5 & 14.8 & -- & -- & 186 $\pm$ 92 & 2.27 $\pm$ 21.84 \\
  0.4 & 170 & 170 & 0.35 $\pm$ 0.02 & -- & -- & 1.8 & 10.8 & 11.5 & $-1.7$ & $-3.3$ & -- & -- \\  \hline
 0.4 & 340 & 460 & 0.41 $\pm$ 0.03 & 229 $\pm$ 194 & 290 $\pm$ 264 & 1.4 & 8.2 & 10.5 & -- & -- & 260 $\pm$ 184 & 0.97 $\pm$ 15.80\\ 
  0.4 & 340 & 340 & 0.39 $\pm$ 0.02 & -- & -- & 1.3 & 5.7 & 6.5 & $-2.5$ & $-4$ & -- & -- \\  \hline
 0.4 & 680 & 920 & 0.41 $\pm$ 0.02 & 640 $\pm$ 296 & 1299 $\pm$ 459 & 1.2 & 5.3 & 7.6 & -- & -- & 970 $\pm$ 273 & 23.55 $\pm$ 61.85 \\
  0.4 & 680 & 680 & 0.39 $\pm$ 0.02 & -- & -- & 1.7 & 9.9 & 10.7 & 4.6 & 3.1 & -- & -- \\ \hline

\multicolumn{13}{c}{z=1.5} \\ \hline
0.4 & 170 & 230 & 0.43 $\pm$ 0.02 & 167 $\pm$ 60 & 266 $\pm$ 115 & 1.1 & 4.6 & 6.9 & -- & -- & 217 $\pm$ 65 & 7.17 $\pm$ 28.91 \\
  0.4 & 170 & 170 & 0.39 $\pm$ 0.02 & -- & -- & 1.5 & 7.5 & 8.3 & $-2.9$ & $-1.4$ & -- & -- \\  \hline
 0.4 & 340 & 460 & 0.39 $\pm$ 0.02 & 410 $\pm$ 253 & 566 $\pm$ 350 & 1.3 & 6.3 & 8.6 & -- & -- & 488 $\pm$ 216 & 2.45 $\pm$ 20.61\\ 
  0.4 & 340 & 340 & 0.38 $\pm$ 0.02 & -- & -- & 1.2 & 3.8 & 4.6 & $-2.5$ & $-4$ & -- & -- \\  \hline
 0.4 & 680 & 920 & 0.39 $\pm$ 0.02 & 2274 $\pm$ 3025 & 3772 $\pm$ 9438 & 1.4 & 8.5 & 10.8 & -- & -- & 3023 $\pm$ 4955 & 9.12 $\pm$ 186.61 \\
  0.4 & 680 & 680 & 0.38 $\pm$ 0.02 & -- & -- & 1.4 & 6.4 & 7.2 & $-2.1$ & $-3.6$ & -- & -- \\ \hline

\multicolumn{13}{c}{z=2} \\ \hline
 0.4 & 170 & 230 & 0.39 $\pm$ 0.03 & 110 $\pm$ 100 & 138 $\pm$ 131 & 1.8 & 12.3 & 14.6 & -- & -- & 124 $\pm$ 82 & 0.86 $\pm$ 15.34\\ 
  0.4 & 170 & 170 & 0.37 $\pm$ 0.02 & -- & -- & 1.7 & 9.7 & 10.4 & $-2.5$ & $-4.2$ & -- & -- \\  \hline
 0.4 & 340 & 460 & 0.38 $\pm$ 0.02 & 233 $\pm$   246 & 283 $\pm$ 339 & 1.5 & 9.6 & 11.9 & -- & -- & 258 $\pm$ 209 & 0.55 $\pm$ 13.78 \\ 
  0.4 & 340 & 340 & 0.37 $\pm$ 0.02 & -- & -- & 1.4 & 6.2 & 6.9 & $-3.4$ & $-5$ & -- & -- \\  \hline
 0.4 & 680 & 920 & 0.39 $\pm$ 0.02 & 1134 $\pm$ 1410 & 1772 $\pm$ 2726 & 1.1 & 4.7 & 7 & -- & -- & 1453 $\pm$ 1535 & 6.35 $\pm$ 93.81 \\
  0.4 & 680 & 680 & 0.38 $\pm$ 0.02 & -- & -- & 1.3 & 4.7 & 5.4 & 0 & $-1.6$ & -- & -- \\ 
 \hline
 \hline

 \multicolumn{3}{c}{}\\

\end{tabular}
\caption{
\textcolor{black}{ Fitting results of data simulated for  model B with errors corresponding 5\% in the measured flux and the assumed redshift z of the source. For the particular mock data each row shows the assumed values of the simulation and the corresponding fitting results obtained for a given redshift. Additionally we fix $M_{1}$ to $ 10^9 M_{\odot}$, $i$ to 35 $\deg$, $R_{\rm{inner}}$ to 6$R_{\rm{g}}$ and $R_{\rm{outer}}$ to 5000 $R_{\rm{g}}$ unless stated otherwise.  We judge and compare the quality of the fit using the reduced chi-square test value $\chi^2_{\nu}$, Akaike \textit{AIC} and Bayesian Information Criteria \textit{BIC} in the corresponding columns. In the column inferred values we provide the calculated values of $d$ and $M_{2}$ of the perturber when possible.}}\label{synth_fit_B}
\end{table*}

\begin{table*}
\centering
\begin{tabular}{ c|c|c|c|c|c|c|c|c|c|c|c|c }

\hline
 \hline 
 
    \multicolumn{3}{c|}{assumed values}& \multicolumn{8}{c|}{fit values} & \multicolumn{2}{c}{inferred values}\\ \hline
  \makecell{$\dot{m}$\\ } & \makecell{$R_{\rm{gap\,in}}$\\ ($R_{\rm{g}}$)} & \makecell{$R_{\rm{gap\,out}}$\\ ($R_{\rm{g}}$)} & {\makecell{$\dot{m}$\\ }}  & \makecell{$R_{\rm{gap\,in}}$\\ ($R_{\rm{g}}$)} & \makecell{$R_{\rm{gap\,out}}$\\ ($R_{\rm{g}}$)} & $\chi^2_{\nu}$ & \textit{AIC} & \textit{BIC} & $\Delta$\,\textit{AIC} & $\Delta$\,\textit{BIC} & \makecell{$d$\\ ($R_{\rm{g}}$)} & \makecell{$M_{2}$\\ ($10^7 M_{\odot}$)}  \\
 \hline

\multicolumn{13}{c}{z=0} \\ \hline
 0.4 & 170 & 230 &  0.42 $\pm$ 0.01  & 143 $\pm$ 11 & 212 $\pm$ 16 & 1.4 & 7.6 & 9.9 & -- & -- & 178 $\pm$ 10 & 4.41 $\pm$ 3.79\\ 
  0.4 & 170 & 170 & 0.35 $\pm$ 0.01 & -- & -- & 4.9 & 26.4 & 27.2 & 18.8 & 17.3 & -- & -- \\  \hline
 0.4 & 340 & 460 & 0.41 $\pm$ 0.01 & 315 $\pm$ 33   & 436 $\pm$ 42 & 1.1 & 3.5 & 5.8 & -- & -- & 376 $\pm$ 27 & 2.51 $\pm$ 3.37 \\ 
  0.4 & 340 & 340 & 0.36 $\pm$ 0.01 & -- & -- & 6.1 & 30 & 30.8 &26.5  & 25 & -- & -- \\  \hline
 0.4 & 680 & 920 & 0.39 $\pm$ 0.01 & 768 $\pm$ 108 & 998 $\pm$ 151 & 1.3 & 7 & 9.3 & -- & -- & 883 $\pm$ 93 & 1.33 $\pm$ 3.24  \\
  0.4 & 680 & 680 & 0.37 $\pm$ 0.01 & -- & -- & 4.1 & 23.6 & 24.4 & 16.6 & 15.1 & -- & -- \\\hline

\multicolumn{13}{c}{z=0.5} \\ \hline
 0.4 & 170 & 230 & 0.40 $\pm$ 0.01 & 146 $\pm$ 17 & 194 $\pm$ 20 & 1.2 & 5.4 & 7.8 & -- & -- & 170 $\pm$ 13 & 1.69 $\pm$ 2.80 \\ 
  0.4 & 170 & 170 & 0.36 $\pm$ 0.01 & -- & -- & 4.8 & 26.1 & 26.9 & 20.7 & 19.1 & -- & -- \\  \hline
 0.4 & 340 & 460 & 0.40 $\pm$ 0.01 & 306 $\pm$ 36   & 408 $\pm$ 52 & 1.1 & 3.5 & 5.9 & -- & -- & 357 $\pm$ 32 & 1.75 $\pm$ 3.29 \\ 
  0.4 & 340 & 340 & 0.37 $\pm$ 0.01 & -- & -- & 5.5 & 28.3 & 29.1 & 24.8 & 23.2 & -- & -- \\  \hline
 0.4 & 680 & 920 & 0.400 $\pm$ 0.005 & 761 $\pm$ 93 & 1085 $\pm$ 128 & 1.5 & 9.6 & 11.9 & -- & -- & 923 $\pm$ 79 & 3.24 $\pm$ 4.83 \\
  0.4 & 680 & 680 & 0.39 $\pm$ 0.01 & -- & -- & 5 & 26.9 & 27.6 & 17.3 & 15.7 & -- & -- \\ \hline

\multicolumn{13}{c}{z=1} \\ \hline
 0.4 & 170 & 230 & 0.39 $\pm$ 0.01 & 170 $\pm$ 20 & 230 $\pm$ 27 & 1.2 & 5.9 & 8.2 & -- & -- & 200 $\pm$ 17 & 2.03 $\pm$ 3.44\\ 
  0.4 & 170 & 170 & 0.36 $\pm$ 0.01 & -- & -- & 6.8 & 31.8 & 32.5 & 25.9 & 24.3 & -- & -- \\  \hline
 0.4 & 340 & 460 & 0.400 $\pm$ 0.004 & 308 $\pm$ 34  & 424 $\pm$ 56 & 1.1 & 4.4 & 6.8 & -- & -- & 366 $\pm$ 33 & 2.39 $\pm$ 4.10 \\ 
  0.4 & 340 & 340 & 0.38 $\pm$ 0.01 & --& --& 5.3 & 27.6 & 28.4 & 23.2 & 21.6 & -- & -- \\  \hline
 0.4 & 680 & 920 & 0.394 $\pm$ 0.004 & 655  $\pm$ 102 & 888 $\pm$ 125 & 1.2 & 4.9 & 7.3 & -- & -- & 772 $\pm$ 81 & 2.10 $\pm$ 4.34 \\
  0.4 & 680 & 680 & 0.38 $\pm$ 0.01 & -- & -- & 3.6 & 21.3 & 22.1 & 16.4 & 14.8 & -- & -- \\ \hline

\multicolumn{13}{c}{z=1.5} \\ \hline
 0.4 & 170 & 230 & 0.40 $\pm$ 0.01 & 144 $\pm$ 18 & 194 $\pm$ 26 & 1.4 & 8.2 & 10.5 & -- & -- & 169 $\pm$ 16 & 1.94 $\pm$ 3.73 \\ 
  0.4 & 170 & 170 & 0.37 $\pm$ 0.01 & -- & -- & 6.9 & 32 & 32.7 & 23.8 & 22.2 & -- & -- \\  \hline
 0.4 & 340 & 460 & 0.393 $\pm$ 0.004 & 344 $\pm$ 42   & 472 $\pm$ 65 & 1.4 & 8.4 & 10.8 & -- & -- & 408 $\pm$ 39 & 2.32 $\pm$ 4.25 \\ 
  0.4 & 340 & 340 & 0.38 $\pm$ 0.01 & -- & -- & 3.9 & 22.8 & 23.6 & 14.4 & 12.8 & -- & -- \\  \hline
 0.4 & 680 & 920 & 0.398 $\pm$ 0.004 & 738 $\pm$ 265 & 982 $\pm$ 321 & 1.1 & 4.5 & 6.8 & -- & -- & 860 $\pm$ 208 & 1.71 $\pm$ 8.85 \\
  0.4 & 680 & 680 & 0.39 $\pm$ 0.01 & -- & -- & 3.9 & 22.9 & 23.6 & 18.4 & 16.8 & -- & -- \\ \hline

\multicolumn{13}{c}{z=2} \\ \hline
 0.4 & 170 & 230 & 0.398 $\pm$ 0.004 & 161  $\pm$ 23 & 207 $\pm$ 34 & 1.2 & 5.3 & 7.6 & -- & -- & 184 $\pm$ 21 & 1.17 $\pm$ 3.16 \\ 
  0.4 & 170 & 170 & 0.38 $\pm$ 0.01 & -- & -- & 4.4 & 24.8 & 25.6 & 19.5 & 18 & -- & -- \\  \hline
 0.4 & 340 & 460 & 0.395 $\pm$ 0.004 & 383 $\pm$ 63   & 502 $\pm$ 78 & 1.1 & 4.8 & 7.1 & -- & -- & 443 $\pm$ 50 & 1.46 $\pm$ 3.72 \\ 
  0.4 & 340 & 340 & 0.388 $\pm$ 0.005 & -- & -- & 2.9 & 17.9 & 18.7 & 13.1 & 11.6 & -- & -- \\  \hline
 0.4 & 680 & 920 & 0.394 $\pm$ 0.003 & 635 $\pm$ 281 & 845 $\pm$ 348 & 1.2 & 5.2 & 7.5 & -- & -- & 740 $\pm$ 224 & 1.71 $\pm$ 11.06 \\
  0.4 & 680 & 680 & 0.39 $\pm$ 0.01 & -- & -- & 4.1 & 23.6 & 24.3 & 18.4 & 16.8 & -- & -- \\ 
 \hline
 \hline

 \multicolumn{3}{c}{}\\

\end{tabular}
\caption{
\textcolor{black}{ Fitting results of data simulated for model B with errors corresponding  1\% in the measured flux. Each row for the particular mock data shows the assumed values of the simulation and fitting results obtained for a given redshift. Additionally we fix $M_{1}$ to $ 10^9 M_{\odot}$, $i$ to 35 $\deg$, $R_{\rm{inner}}$ to 6$R_{\rm{g}}$ and $R_{\rm{outer}}$ to 5000 $R_{\rm{g}}$ unless stated otherwise. The three criteria, i.e., $\chi^2_{\nu}$, Akaike \textit{AIC} and Bayesian \textit{BIC} provide a mutually consistent estimation of the fit goodness. In the column of `inferred values' we give the calculated values of $d$ and $M_{2}$ of the perturber when possible. Compared to the previous table, now a smaller level of the assumed error allows us to recover the parameters from the mock data very well (see the main text for further details). }}\label{synth_fit_B2}
\end{table*}

\section{Discussion}
\label{sec_discussion}
We investigated the effect of the ADAF as well as the gap created due to an orbiting secondary body on the SED of a standard thin disc that is present in AGN. Both the presence of the ADAF and the second massive black hole are plausible in galactic nuclei during their evolution. In fact, during the final phases of galaxy mergers, i.e. during the binary black hole inspiral, a gap is initially formed at intermediate scales \citep{gultekin}, followed by the hollow formation just before the binary black-hole merger \citep{Shapiro_2010}. However, the coexistence of these ``inspiral'' perturbations with the preexisting inner ADAF or the ADAF region that forms during the inspiral has not been investigated in detail. 
These perturbations are therefore astrophysically relevant and could be traced in real photometric data in case the photometric errors are of the order of \textcolor{black}{ $5\%$}, though higher precision is necessary for a significant detection of smaller gaps at the distance of a few 100 gravitational radii, i.e. at intermediate distance scales. 

\subsection{Limitations of the adopted picture}

\textcolor{black}{ In model A and C setups, we tie the inner radius of the thin disk to the ADAF radius, which represents the inner gap in the thermal disk continuum. For simplicity and a model definition, the ADAF outer radius is set by the accretion rate (and the viscosity parameter). In the adopted specific model, i.e. the strong ADAF principle, the outer radius of the ADAF region as predicted by several theoretical works \citep[see e.g.][]{1995ApJ...438L..37A, 1996PASJ...48...77H, 1998PASJ...50..559K} predicts steep dependence on the accretion rate, $\propto \dot{m}^{-2}$, which is in tension with the SEDs of the standard population of quasars in a sense that it predicts larger ADAF sizes and thus depressions in the UV luminosity than indicated by the observationally inferred accretion rates \citep[see e.g.][]{2020A&A...642A.150L}. More effects may be needed to address the inner regions of accretion flows properly, including also thermal conduction and poloidal magnetic field, which both predict different ADAF radii with a flatter dependence on the accretion rate according to Eqs.~\eqref{eq_evaporation_radius} and \eqref{eq_magnetospheric_radius}, respectively, in the further discussion. An extreme example is the nearby AGN NGC 3147 \citep{2019MNRAS.488L...1B}, where a thin disk exists on the scales where most theories predict a hot flow for the Eddington ratio of $\dot{m}\sim 10^{-4}$ (Eq.~\eqref{r_adaf_eq} predicts the hot flow for the whole disk extent). In contrast, a realistic accretion disc solution for intermediate accretion rates between low-luminous and high-luminous sources is still under investigation. Recently, a stable ``puffy'' disk model \citep{2019ApJ...884L..37L} was proposed for sources accreting at the fraction of the Eddington limit based on radiative GRMHD simulations. The model consists of optically thick thinner disk embedded in geometrically thicker and turbulent warm corona component and a hot diluted outflowing part. Hence, the realistic solutions applicable to the broadband SEDs of quasars are rather complex. Generating broadband spectra from these multicomponent models is therefore computationally challenging, even more when using them for scanning a certain parameter space or fitting them to the quasar SED data. The strong ADAF principle assumed in the draft while generating accretion disk SEDs as a function of the accretion rate is a simplistic approach on one hand, on the other hand, it can be easily applied for scanning a broader parameter space, including the accretion rate, inclination, black hole mass, and the inner radius of the disc. We agree that it does not reproduce the quasar mean spectrum in the full complexity. Still, it may be qualitatively applicable to peculiar AGN sources, which deviate from the standard AGN population, e.g., changing-look AGN. For these and other peculiar sources, the general parametrization with a free inner radius is sometimes required to capture the broad-band SED , see e.g.,  IC 4329A \citep{2021MNRAS.504.4015D} with a wide inner void with the radius of $\sim 80-150\,R_{\rm g}$ for the SMBH mass of $\sim 1-2 \times 10^8\,M_{\odot}$, or NGC 7469 and Mrk 352 that show indications of the inner void at the scales of $\sim 35-125\,R_{\rm g}$ and $\sim 50-135\,R_{\rm g}$ \citep{2023arXiv230311882K}, respectively, while fixing the inner radius at the ISCO would be in tension with the observed spectra.}

\textcolor{black}{ The simplistic scenario of eq. (\ref{gap_luminostiy}) will need further amendments, which may be relevant for an AGN with a secondary SMBH that is massive enough to clear a gap within the accretion disk. It has been based on the scenario of a standard, geometrically thin accretion disk, where any mass element that crosses the innermost stable circular radius becomes rapidly accreted on the free-fall
time-scale. In fact, it has been recognized  \citep[][and further references cited therein]{1976A&A....53..267S,1997ApJ...488..109R} that the no-torque condition at ISCO may be violated, for example, by the effect of magnetic fields, but the actual details of the mechanism and the resulting spectral signatures consequences to the disk structure are being investigated in the literature to date. In even more complicated case of a radio loud (jetted) primary SMBH, the influence of the gap on an outgoing SED can be further enhanced due to Compton upscattering of low-energy photons from the gappy disk on the non-thermal radiation of the jet (this interesting situation is considered by \citeauthor{2016A&A...588A.125R} \citeyear{2016A&A...588A.125R}). We postpone the investigation of an enhanced effect of jet upon the emerging SED from a gappy accretion disk to a future work but we note that the effects on the spectrum are then expected to be more pronounced.} 

Once we assume the disc perturbation in the form of an ADAF region, a gap due to an orbiting secondary body, or the combination of both, the SED flux density is suppressed when compared to the unperturbed case. The SED simulations corresponding to model A show strong effects of the ADAF region in high-energy bands (see Figure \ref{fig:adaf.png}), which is also visible in the optical range if the accretion disc/ADAF transition radius is of the order of tens of gravitational radii. This kind of model can effectively be employed when setting the constraints for the size of the inner region of the accretion disc \citep[e.g.][]{2020ApJ...896L..36Z,  2021MNRAS.504.4015D,2023arXiv230311882K}. The SED spectra in model B scenario, however, are only weakly suppressed, forming a shallow feature, and only larger gaps can be
detected in the SED (see Figures \ref{fig:sec_BH_revised_masses} and \ref{sec_BH_revised}). We observe that for the \textcolor{black}{ $5\%$} errors, model A scenario can be clearly distinguished by the means of SED spectral changes \textcolor{black}{ (see Table \ref{synth_fit_A})}. \textcolor{black}{ The situation does not change if we introduce a dust component in model A (see Table \ref{synth_fit_dust}) and we show it is possible to differentiate the inner ADAF from the intrinsic reddening in the source. Using the model A with or without the dust component for fitting our mock data, we recover better results for higher values of the redshift as the spectrum is then covering the frequency range with most noticeable changes.
However, the perturbation due to a secondary body introduced in model B ($\lesssim 12\%$ change in the spectra for the ratio of the secondary mass to the primary mass $\leq 10^{-2}$) cannot be distinguished from the unperturbed case when assuming 5\% errors \textcolor{black}{ (see Table \ref{synth_fit_B}}). We repeat the fitting with the exact same mock data having 1\% errors and show using the $AIC$ and $BIC$ ($\Delta AIC$ and $\Delta BIC$, respectively) that the perturbed model is favoured (see Table \ref{synth_fit_B2}). The effect of the redshift is now reversed as for the model A. The SED changes due to the secondary component are mostly in the low frequency range and therefore fitting the mock data in the higher redshift regime does not lead to reasonable values. We are also able to infer the mass of the secondary and its distance using the fit results from Table \ref{synth_fit_B} and \ref{synth_fit_B2} and see a major improvement in precision as we switch to 1\% errors.  }

\subsection{Components of the toy model}
\textcolor{black}{ Let us remind the reader that the approach adopted in this paper assumes the model setup consisting of the individual components (i.e., sections of a standard accretion disc separated by gaps due to an embedded secondary orbiter and the inner ADAF region). Based on these assumptions we checked whether the model parameters can be constrained. Quite unsurprisingly, the opposite approach (starting from the continuum spectral profile) would not be unique enough to determine the model geometry; this may change when the line spectra are taken into account (work in progress).} 

Changing the total size of the accretion disc could prove to help the perturbation to manifest in a cleaner way in simulations. However, as the distance of the secondary body is of the order of 100 $R_{\rm{g}}$, the effects on the SED will still be located in the low-frequency band, at the border between the optical and the near-IR bands. Accepting the presence of the secondary component below 80 $R_{\rm{g}}$ for the primary mass of $M_1=10^9\,M_{\odot}$, i.e. assuming the merger timescale smaller than $t_{\rm{merge}} \lesssim 10^4 {\rm yr}$, would shift the dip in the SED to a higher frequency or allow us to study the case of a broad gap located in the inner disc region below $\sim 100R_{\rm{g}}$ or spanning down to it \citep[see e.g.][]{gultekin}.

Focusing on the combination of the latter two perturbation scenarios, we construct model C and set constraints on the frequency range where the spectral changes are most significant (see Figure \ref{model_C_comparison}). We notice that SED simulation results for model C show that the size of the ADAF region can extend up to the secondary gap, or further, depending on the value of the relative accretion rate. In another words, with a relative accretion rate of the order of 0.1, the effect of the secondary perturber will not be captured since the gap opened by an orbiting perturber gets engulfed by the extended ADAF region (see Figure \ref{adaf+sec_BH_revised}).

The distance of the secondary component from the primary one (model B and C scenarios) of the order of a few $100\,R_{\rm g}$ was adopted based on the time-scale on which the
    two black holes are expected to merge due to the emission of gravitational waves. A standard approximation for the circular two-body problem gives \citep{1964PhRv..136.1224P,shapiro_and_teukolsky_1983}

    \begin{align}
    t_{\rm merge} &=\frac{5c^5}{256 G^3}\frac{a_0^4}{M_1 M_2 (M_1+M_2)}\,\notag\\
    & = \frac{5c^5}{256 G^3} \frac{a_0^4}{x_{\rm p}x_{\rm s}M_{\rm tot}^3}\,\notag\\
    & \approx 12\,500 \left(\frac{a_0}{80\,R_{\rm g}} \right)^4 \left(\frac{x_{\rm p}}{1} \right)^{-1} \left(\frac{x_{\rm s}}{10^{-2}} \right)^{-1} \left(\frac{M_{\rm tot}}{10^9\,M_{\odot}} \right)^{-3}\,{\rm yr}\,,
        \label{t_merge}
    \end{align}
    where $a_0$ is the initial orbital distance of the secondary component with respect to the primary one, $M_{\rm{tot}}=M_1+M_2$ is the total mass of the system, $x_{\rm{p}}=M_1/M_{\rm tot}$ and $x_{\rm{s}}=M_2/M_{\rm tot}$ are the ratios of the mass of primary component and secondary component to the total mass of the system, respectively. Considering $t_{\rm merge}\gtrsim 10^4\,{\rm yr}$, we adopted the distances $d \gtrsim 100 R_{\rm{g}}$ to ensure the overall stability of the two-body system and hence the quasi-stationarity of the accretion flow, which implies the gap width of $\Delta R\gtrsim 30\,R_{\rm g}$ according to Eq.~\eqref{distance_gap_size}.

We modeled the position of the inner ADAF assuming the \textit{strong ADAF principle} (see eq. (\ref{r_adaf_eq})).  An alternative, physically motivated model for the transition from the cold accretion disc to the inner ADAF region would be the evaporation of the cold disc due to the electron conduction between the cold disc and the two-temperature hot corona \citep{2000A&A...360.1170R,2002A&A...392L...5M}. The relation for the evaporation radius according to \citet{2004A&A...428...39C}  takes into account the magnetic pressure and is consistent with the extension of the broad-line region,
    \begin{equation}
        R_{\rm evap}=39 \alpha_{0.1}^{0.8} \beta^{-1.08} \dot{m}^{-0.53}\,R_{\rm g}\,,
        \label{eq_evaporation_radius}
    \end{equation}
        where $\beta$ is the ratio between the total (gas $+$ radiation pressure) and the total and the magnetic pressure. Hence, it takes the value between zero for the magnetically-dominated case and one for a negligible magnetic field. In comparison with Eq.~\eqref{r_adaf_eq} relevant for the ADAF principle, the evaporation of the cold disc due to electron conduction can in principle result in the extended inner hot flow even for the accretion rate close to the Eddington limit.  Alternatively, the formation of the magnetically arrested disc (MAD) can open a large inner gap or a hole under the assumption that the accretion drags inside the magnetic flux frozen in plasma \citep{1974Ap&SS..28...45B,1976Ap&SS..42..401B}. When the sufficient amount of the poloidal magnetic flux is accumulated, the axisymmetric accretion disc becomes unstable and breaks into magnetically confined islands that move inwards slower than the free-fall velocity via magnetic reconnection and interchanges. The transition occurs at the magnetospheric radius that can be estimated for the axisymmetric flow as follows \citep{2003PASJ...55L..69N,2020ApJ...897...99T},
        \begin{align}
           R_{\rm m}&=\left(\frac{\dot{M}\sqrt{GM_{\bullet}}}{2^{1/2}\epsilon B_{\rm pol}^2} \right)^{2/5} \,\notag\\
           &\sim 48.5\,\left(\frac{M_{\bullet}}{10^9\,M_{\odot}} \right)^{1/5}\left(\frac{\dot{m}}{0.1} \right)^{2/5}\left(\frac{B_{\rm pol}}{10^3\,{\rm G}} \right)^{-4/5}\left(\frac{\epsilon}{10^{-2}} \right)^{-2/5}R_{\rm g}\,,
           \label{eq_magnetospheric_radius}
        \end{align}
        where $\epsilon$ expresses the ratio between the radial flow velocity below $R_{\rm m}$ and the free-fall velocity and is typically $\epsilon\sim 10^{-3}-10^{-2}$ \citep{2003PASJ...55L..69N}, and $B_{\rm pol}$ is the poloidal magnetic field strength.
        Both the formation of ADAF due to evaporation and the MAD state give a different relation between the accretion rate and the transition radius than adopted in the current paper, see Eq.~\eqref{r_adaf_eq} for comparison, hence more free parameters would have to be adopted in simulations to properly describe these processes. There are also alternative scenarios for the drop detection in the far-UV/soft X-ray domain. For instance, the ionized dust-free absorber partially covering the central source is consistent with the SED flux decrease in the high-energy band of Mrk 817 \citep{kara2021agn}. 

As for the secondary gap component, recent MHD numerical simulations demonstrate another way of gap formation via the supernova explosion in the accretion disc \textcolor{black}{ \citep[e.g.][]{2021ApJ...906...15M, 2023ApJ...944..217R}}. 
However, in this case one needs to consider that the supernova explosion sweeps away the material on much larger spatial scales and can create a gap with the size bigger than the ones proposed in our article, even the whole inner region, particularly for the black-hole mass smaller than $10^8 M_{\odot}$. The possibility of multiple objects such as stars and compact remnants accumulating in so-called ``migration traps'' could act as a substitute for a secondary gap component as well \citep[e.g.][]{2016ApJ...819L..17B, 2022ApJ...926..101M}. The stellar-sized objects can accumulate at a few hundred gravitational radii since the migration torque points not only inwards but also outwards. It is also a possibility that several gaps in the disc are created due to the orbital resonances of the gas with respect to the orbiting secondary massive body, i.e. in a way analogous to the formation of Kirkwood gaps in the Main Asteroid Belt in the Sun-Jupiter system. 

\subsection{Truncated accretion disk and emergence of ADAF region}
\cite{below_adaf} show that there is also a possibility that the ADAF region does not extend down to the ISCO. Instead, a cold slab of gas develops below it due to the ADAF recondensation. Such a scenario would lead to the superposition of radiation from two sub-discs (assuming model A) or three sub-discs (model C), and therefore it would result in a qualitatively different SED. Due to additional parameters necessary for the description of the recondensation region, we do not consider this scenario in the current study.

The model of the standard disc with gaps may be an attractive scenario for quasars where periodic changes in continuum or emission lines are claimed to be detected  \citep[e.g.][and the references therein]{liu2019,chen2020,chen2022}. By now there are 250 such sources, but the false detection rate can be high \citep[e.g.][]{witt2021}. \citet{Guo_2019} argued that the presence of the secondary companion should lead to the departure from the standard accretion disc models.
Thus, our model can be applied to quasar samples with a good photometric or spectroscopic coverage (in the latter case the contribution of the emission features should be subtracted as the model aims at modelling continuum only). It could provide an independent method from periodicity searches to identify the binary black hole candidates. 

To make a conservative estimate of the number of ``gappy'' AGN, i.e. those likely having two gaps, in the local Universe, we consider the volume density of Milky Way-like galaxies, $\Phi_{\rm MW}\approx 0.006\,{\rm Mpc^{-3}}$, which follows from the Schechter luminosity function \citep{2015eaci.book.....S}. When we consider the total number of AGN up to the intermediate redshift of $z=0.5$, we get $N_{\rm AGN}=\Phi_{\rm MW}V_{\rm C}(z)f_{\rm AGN}\approx 1.7\times 10^7$, where $V_{\rm C}$ is the corresponding comoving volume at the redshift of $0.5$ and we set the fraction of AGN to $f_{\rm AGN}\sim 0.1$. Subsequently, the fraction of type 1 AGN up to the viewing angle of $40^{\circ}$ is $f_{\rm I}\approx 0.64$ if we consider the uniform distribution of viewing angle cosines up to 90 degrees. The merger fraction of massive galaxies is taken as $f_{\rm merge}\sim 0.1$ with the frequency $\nu_{\rm merge}=1\,{\rm Gyr}^{-1}$ \citep{2021MNRAS.501.3215O}. If we consider the timescale of $\tau_{\rm merge}\sim 10^4\,{\rm yr}$ of the last stage of the merger capable of creating the gap in the disc at intermediate distances, we get the number of these potentially gappy AGN $N_{\rm gap}=\Phi_{\rm MW}V_{\rm C}(z)f_{\rm AGN}f_{\rm I}f_{\rm merge}\nu_{\rm merge} \tau_{\rm merge}\approx 11$. Another estimate comes from the extended Press--Schechter formalism for merging halos. According to \citet{2006MNRAS.371.1992E}, the rate of SMBH mergers per comoving volume at $z=0.5$ for SMBHs more massive than $10^5\,M_{\odot}$ is $\eta_{\rm merge}\approx 2\times 10^{-12}\,{\rm Mpc^{-3}\,yr^{-1}}$. The estimated total number of SMBH--SMBH merging systems up to $\tau_{\rm merge}=10^4$ years is $N_{\rm gap}'\sim \eta_{\rm merge}V_{\rm C}(z)\tau_{\rm merge}\sim 535$. Hence, we expect of the order of $10$--$100$ systems with perturbed AGN discs due to mergers from the local Universe up to the intermediate redshifts. The perturbation due to galaxy mergers takes place in addition to inner ADAFs, which are generally expected to occur in sub-Eddington systems. Overall, considering a large redshift range up to $z\approx 6$, the Legacy Survey of Space and Time (LSST) that will be performed by the Vera C. Rubin Observatory \citep{2019ApJ...873..111I} will discover tens of millions of quasars \citep{2018arXiv181106542B}. A large fraction of those will be binaries resulting from galaxy mergers that are thought to trigger AGN activity. \citet{2021MNRAS.506.2408X} predict the detection of $\sim 20-100$ SMBH binaries in the mass range $10^5-10^9\,M_{\odot}$, the redshift range of $0-6$, and the orbital periods of $1-70$ days. Out of those, $10-150$ sources could be ultracompact binaries with orbital periods of $\lesssim 1$ day. This order of magnitude is also consistent with the number of self-lensing binaries that could be detected by the Vera C. Rubin observatory \citep{2021MNRAS.508.2524K}. Hence, the Vera C. Rubin LSST observatory will provide candidate sources for follow-up high-precision broad-band photometry to explore potential signs of perturbed accretion discs.

There are two apparent limitations of the model. First, the current accretion disc models considered in this study do not include the effect of the spin. Thus, we must note that if the inner ADAF is rather compact, a counter-rotating disc can be misidentified as an ADAF. However, if the outer ADAF radius (inner radius of the standard disc) is larger than 10$R_g$ relativistic effects related to the spin are less important. In any case, detecting radii of the inner ADAF of the order of ten gravitational radii require well-sampled far-UV data and/or a high redshift of the source. The UV domain of the nearby bright AGN can be detected with the sufficient precision of $\sim 0.005$ mag with relatively small space UV telescopes, see e.g. \citet{2022arXiv220705485W}.

\section{Conclusions}
\label{sec_conclusions}

We studied the possibility to model the broad-band SED when the multi-color thermal component from the standard Shakura-Sunyaev accretion disc is perturbed. The proposed models include a nearly spherical ADAF component in the innermost region (representing the hot corona), a secondary body (presumably, an intermediate or supermassive black hole) that clears a gap in the standard disc, or their combination (marked as exemplary models A, B and C, respectively). \textcolor{black}{ We also considered the effect of dust extinction to demonstrate that the inner ADAF region can be distinguished from the intrinsic reddening in the source. The set of three exemplary cases captures qualitatively different situations for different redshift values. Despite that the adopted model is very speculative, it can help us grasp the effects that are expected in a widely discussed scenario of an accreting SMBH with a secondary, i.e. a less massive orbiter embedded in the standard accretion flow that can be intrinsically truncated at a certain radius. The selected models share a common attribute: the perturbation of the standard-disc scenario diminishes a section of the emerging SED with respect to the base (unperturbed) accretion disc. Reconstructing the parameters of the inner ADAF can be achieved with a relatively high accuracy if the data covers the rest-frame far-UV domain, and the ADAF size is extended, i.e. of the order of tens gravitational radii. Under favourable conditions, the photometric data with the uncertainties of the order of \textcolor{black}{ $5\%$} would be sufficient to recover the SED changes caused by the presence of an ADAF region. On the other hand, the determination of the parameters of the gap (model B) requires data with small errors at the level of \textcolor{black}{ $1\%$}, since the SED suppression is only very shallow in this case.}

In summary, in this paper we argued that the gappy accretion disc with an embedded perturber near a supermassive black hole can offer an explanation of certain deviations from the standard model that have been noticed in SEDs of various quasars and AGN in the optical/UV to soft X-ray domains, such as the departures customarily parameterized by $f_{\rm col}$. Our tentative interpretation in terms of the effect of gaps is an interesting (albeit not unique) possibility. A detailed fitting to actual observational data and exploring the parameter degeneracies will be necessary but it remains beyond the scope of the present paper; this work is in progress.
    
\section*{Acknowledgements}
The authors thank an anonymous referee for very useful suggestions that helped us to expand our modelling and clarify several important points in the first version of the manuscript. M\v{S} and VK acknowledge support from the Czech Science Foundation research grant No. 21-11268S ``Mass and charge currents in general relativity and astrophysics'', and the Czech Academy of Sciences project CONICET-22-06 titled ``Accretion onto black holes on different scales: outflows and interactions''. MZ acknowledges  GA\v{C}R EXPRO grant (No. 21-13491X) ``Exploring the Hot Universe and Understanding Cosmic Feedback''. 

\section*{Data Availability}
 \textcolor{black}{ The synthetic data underlying this article will be shared upon reasonable request to the corresponding author (M\v{S}).}




\bibliographystyle{mnras}
\bibliography{example} 

\begin{thebibliography}{}
\makeatletter
\relax
\def\mn@urlcharsother{\let\do\@makeother \do\$\do\&\do\#\do\^\do\_\do\%\do\~}
\def\mn@doi{\begingroup\mn@urlcharsother \@ifnextchar [ {\mn@doi@}
  {\mn@doi@[]}}
\def\mn@doi@[#1]#2{\def\@tempa{#1}\ifx\@tempa\@empty \href
  {http://dx.doi.org/#2} {doi:#2}\else \href {http://dx.doi.org/#2} {#1}\fi
  \endgroup}
\def\mn@eprint#1#2{\mn@eprint@#1:#2::\@nil}
\def\mn@eprint@arXiv#1{\href {http://arxiv.org/abs/#1} {{\tt arXiv:#1}}}
\def\mn@eprint@dblp#1{\href {http://dblp.uni-trier.de/rec/bibtex/#1.xml}
  {dblp:#1}}
\def\mn@eprint@#1:#2:#3:#4\@nil{\def\@tempa {#1}\def\@tempb {#2}\def\@tempc
  {#3}\ifx \@tempc \@empty \let \@tempc \@tempb \let \@tempb \@tempa \fi \ifx
  \@tempb \@empty \def\@tempb {arXiv}\fi \@ifundefined
  {mn@eprint@\@tempb}{\@tempb:\@tempc}{\expandafter \expandafter \csname
  mn@eprint@\@tempb\endcsname \expandafter{\@tempc}}}

\bibitem[\protect\citeauthoryear{{Abramowicz}, {Czerny}, {Lasota}  \&
  {Szuszkiewicz}}{{Abramowicz} et~al.}{1988}]{abramowicz88}
{Abramowicz} M.~A.,  {Czerny} B.,  {Lasota} J.~P.,   {Szuszkiewicz} E.,  1988,
  \mn@doi [\apj] {10.1086/166683}, \href
  {https://ui.adsabs.harvard.edu/abs/1988ApJ...332..646A} {332, 646}

\bibitem[\protect\citeauthoryear{{Abramowicz}, {Chen}, {Kato}, {Lasota}  \&
  {Regev}}{{Abramowicz} et~al.}{1995}]{1995ApJ...438L..37A}
{Abramowicz} M.~A.,  {Chen} X.,  {Kato} S.,  {Lasota} J.-P.,   {Regev} O.,
  1995, \mn@doi [\apjl] {10.1086/187709}, \href
  {https://ui.adsabs.harvard.edu/abs/1995ApJ...438L..37A} {438, L37}

\bibitem[\protect\citeauthoryear{{Abramowicz}, {Chen}, {Granath}  \&
  {Lasota}}{{Abramowicz} et~al.}{1996}]{1996ApJ...471..762A}
{Abramowicz} M.~A.,  {Chen} X.~M.,  {Granath} M.,   {Lasota} J.~P.,  1996,
  \mn@doi [\apj] {10.1086/178004}, \href
  {https://ui.adsabs.harvard.edu/abs/1996ApJ...471..762A} {471, 762}

\bibitem[\protect\citeauthoryear{Akaike}{Akaike}{1973}]{akaike1973information}
Akaike H.,  1973, Information Theory and an Extension of the Maximum Likelihood
  Principle.
Springer New York, New York, NY, pp 199--213

\bibitem[\protect\citeauthoryear{{Alexander}}{{Alexander}}{2017}]{2017ARA&A..55...17A}
{Alexander} T.,  2017, \mn@doi [\araa] {10.1146/annurev-astro-091916-055306},
  \href {https://ui.adsabs.harvard.edu/abs/2017ARA&A..55...17A} {55, 17}

\bibitem[\protect\citeauthoryear{{Balbus} \& {Hawley}}{{Balbus} \&
  {Hawley}}{1991}]{balbus_MRI}
{Balbus} S.~A.,  {Hawley} J.~F.,  1991, \mn@doi [\apj] {10.1086/170270}, \href
  {https://ui.adsabs.harvard.edu/abs/1991ApJ...376..214B} {376, 214}

\bibitem[\protect\citeauthoryear{{Balbus} \& {Mummery}}{{Balbus} \&
  {Mummery}}{2018}]{2018MNRAS.481.3348B}
{Balbus} S.~A.,  {Mummery} A.,  2018, \mn@doi [\mnras] {10.1093/mnras/sty2467},
  \href {https://ui.adsabs.harvard.edu/abs/2018MNRAS.481.3348B} {481, 3348}

\bibitem[\protect\citeauthoryear{{Bardeen}, {Press}  \& {Teukolsky}}{{Bardeen}
  et~al.}{1972}]{1972ApJ...178..347B}
{Bardeen} J.~M.,  {Press} W.~H.,   {Teukolsky} S.~A.,  1972, \mn@doi [\apj]
  {10.1086/151796}, \href
  {https://ui.adsabs.harvard.edu/abs/1972ApJ...178..347B} {178, 347}

\bibitem[\protect\citeauthoryear{{Bardeen}, {Carter}  \& {Hawking}}{{Bardeen}
  et~al.}{1973}]{1973CMaPh..31..161B}
{Bardeen} J.~M.,  {Carter} B.,   {Hawking} S.~W.,  1973, \mn@doi
  [Communications in Mathematical Physics] {10.1007/BF01645742}, \href
  {https://ui.adsabs.harvard.edu/abs/1973CMaPh..31..161B} {31, 161}

\bibitem[\protect\citeauthoryear{{Bellovary}, {Mac Low}, {McKernan}  \&
  {Ford}}{{Bellovary} et~al.}{2016}]{2016ApJ...819L..17B}
{Bellovary} J.~M.,  {Mac Low} M.-M.,  {McKernan} B.,   {Ford} K.~E.~S.,  2016,
  \mn@doi [\apjl] {10.3847/2041-8205/819/2/L17}, \href
  {https://ui.adsabs.harvard.edu/abs/2016ApJ...819L..17B} {819, L17}

\bibitem[\protect\citeauthoryear{{Ben-Ami} et~al.,}{{Ben-Ami}
  et~al.}{2022}]{2022SPIE12181E..05B}
{Ben-Ami} S.,  et~al., 2022, in {den Herder} J.-W.~A.,  {Nikzad} S.,
  {Nakazawa} K.,  eds,  Society of Photo-Optical Instrumentation Engineers
  (SPIE) Conference Series Vol. 12181, Space Telescopes and Instrumentation
  2022: Ultraviolet to Gamma Ray. p. 1218105 (\mn@eprint {arXiv} {2208.00159}),
  \mn@doi{10.1117/12.2629850}

\bibitem[\protect\citeauthoryear{{Bianchi} et~al.,}{{Bianchi}
  et~al.}{2019}]{2019MNRAS.488L...1B}
{Bianchi} S.,  et~al., 2019, \mn@doi [\mnras] {10.1093/mnrasl/slz080}, \href
  {https://ui.adsabs.harvard.edu/abs/2019MNRAS.488L...1B} {488, L1}

\bibitem[\protect\citeauthoryear{{Binette}, {Krongold}, {Haro-Corzo},
  {Humphrey}  \& {Morais}}{{Binette} et~al.}{2022}]{binette2022}
{Binette} L.,  {Krongold} Y.,  {Haro-Corzo} S. A.~R.,  {Humphrey} A.,
  {Morais} S.~G.,  2022, arXiv e-prints, \href
  {https://ui.adsabs.harvard.edu/abs/2022arXiv221104660B} {p. arXiv:2211.04660}

\bibitem[\protect\citeauthoryear{{Bisnovatyi-Kogan} \&
  {Ruzmaikin}}{{Bisnovatyi-Kogan} \& {Ruzmaikin}}{1974}]{1974Ap&SS..28...45B}
{Bisnovatyi-Kogan} G.~S.,  {Ruzmaikin} A.~A.,  1974, \mn@doi [\apss]
  {10.1007/BF00642237}, \href
  {https://ui.adsabs.harvard.edu/abs/1974Ap&SS..28...45B} {28, 45}

\bibitem[\protect\citeauthoryear{{Bisnovatyi-Kogan} \&
  {Ruzmaikin}}{{Bisnovatyi-Kogan} \& {Ruzmaikin}}{1976}]{1976Ap&SS..42..401B}
{Bisnovatyi-Kogan} G.~S.,  {Ruzmaikin} A.~A.,  1976, \mn@doi [\apss]
  {10.1007/BF01225967}, \href
  {https://ui.adsabs.harvard.edu/abs/1976Ap&SS..42..401B} {42, 401}

\bibitem[\protect\citeauthoryear{{Brandt} et~al.,}{{Brandt}
  et~al.}{2018}]{2018arXiv181106542B}
{Brandt} W.~N.,  et~al., 2018, arXiv e-prints, \href
  {https://ui.adsabs.harvard.edu/abs/2018arXiv181106542B} {p. arXiv:1811.06542}

\bibitem[\protect\citeauthoryear{{Calzetti}, {Kinney}  \&
  {Storchi-Bergmann}}{{Calzetti} et~al.}{1994}]{calzetti1994}
{Calzetti} D.,  {Kinney} A.~L.,   {Storchi-Bergmann} T.,  1994, \mn@doi [\apj]
  {10.1086/174346}, \href
  {https://ui.adsabs.harvard.edu/abs/1994ApJ...429..582C} {429, 582}

\bibitem[\protect\citeauthoryear{{Capellupo}, {Netzer}, {Lira}, {Trakhtenbrot}
  \& {Mej{\'\i}a-Restrepo}}{{Capellupo} et~al.}{2015}]{2015MNRAS.446.3427C}
{Capellupo} D.~M.,  {Netzer} H.,  {Lira} P.,  {Trakhtenbrot} B.,
  {Mej{\'\i}a-Restrepo} J.,  2015, \mn@doi [\mnras] {10.1093/mnras/stu2266},
  \href {https://ui.adsabs.harvard.edu/abs/2015MNRAS.446.3427C} {446, 3427}

\bibitem[\protect\citeauthoryear{{Cardelli}, {Clayton}  \& {Mathis}}{{Cardelli}
  et~al.}{1989}]{cardelli1989}
{Cardelli} J.~A.,  {Clayton} G.~C.,   {Mathis} J.~S.,  1989, \mn@doi [\apj]
  {10.1086/167900}, \href
  {https://ui.adsabs.harvard.edu/abs/1989ApJ...345..245C} {345, 245}

\bibitem[\protect\citeauthoryear{{Chakrabarti}}{{Chakrabarti}}{1996}]{1996PhRvD..53.2901C}
{Chakrabarti} S.~K.,  1996, \mn@doi [\prd] {10.1103/PhysRevD.53.2901}, \href
  {https://ui.adsabs.harvard.edu/abs/1996PhRvD..53.2901C} {53, 2901}

\bibitem[\protect\citeauthoryear{{Chen} et~al.,}{{Chen}
  et~al.}{2020}]{chen2020}
{Chen} Y.-C.,  et~al., 2020, \mn@doi [\mnras] {10.1093/mnras/staa2957}, \href
  {https://ui.adsabs.harvard.edu/abs/2020MNRAS.499.2245C} {499, 2245}

\bibitem[\protect\citeauthoryear{{Chen} et~al.,}{{Chen}
  et~al.}{2022}]{chen2022}
{Chen} Y.-J.,  et~al., 2022, \mn@doi [arXiv e-prints]
  {10.48550/arXiv.2206.11497}, \href
  {https://ui.adsabs.harvard.edu/abs/2022arXiv220611497C} {p. arXiv:2206.11497}

\bibitem[\protect\citeauthoryear{{Collier} \& {Peterson}}{{Collier} \&
  {Peterson}}{2001}]{2001ApJ...555..775C}
{Collier} S.,  {Peterson} B.~M.,  2001, \mn@doi [\apj] {10.1086/321517}, \href
  {https://ui.adsabs.harvard.edu/abs/2001ApJ...555..775C} {555, 775}

\bibitem[\protect\citeauthoryear{{Collin} \& {Hur{\'e}}}{{Collin} \&
  {Hur{\'e}}}{1999}]{1999A&A...341..385C}
{Collin} S.,  {Hur{\'e}} J.-M.,  1999, \aap, \href
  {https://ui.adsabs.harvard.edu/abs/1999A&A...341..385C} {341, 385}

\bibitem[\protect\citeauthoryear{{Czerny} \& {Elvis}}{{Czerny} \&
  {Elvis}}{1987}]{1987ApJ...321..305C}
{Czerny} B.,  {Elvis} M.,  1987, \mn@doi [\apj] {10.1086/165630}, \href
  {https://ui.adsabs.harvard.edu/abs/1987ApJ...321..305C} {321, 305}

\bibitem[\protect\citeauthoryear{{Czerny}, {Li}, {Loska}  \&
  {Szczerba}}{{Czerny} et~al.}{2004a}]{czerny_li2004}
{Czerny} B.,  {Li} J.,  {Loska} Z.,   {Szczerba} R.,  2004a, \mn@doi [\mnras]
  {10.1111/j.1365-2966.2004.07590.x}, \href
  {https://ui.adsabs.harvard.edu/abs/2004MNRAS.348L..54C} {348, L54}

\bibitem[\protect\citeauthoryear{{Czerny}, {R{\'o}\.za{\'n}ska}  \&
  {Kuraszkiewicz}}{{Czerny} et~al.}{2004b}]{2004A&A...428...39C}
{Czerny} B.,  {R{\'o}\.za{\'n}ska} A.,   {Kuraszkiewicz} J.,  2004b, \mn@doi
  [\aap] {10.1051/0004-6361:20040487}, \href
  {https://ui.adsabs.harvard.edu/abs/2004A&A...428...39C} {428, 39}

\bibitem[\protect\citeauthoryear{{\lowercase{D}'Ascoli}, {Noble}, {Bowen},
  {Campanelli}, {Krolik}  \& {Mewes}}{{\lowercase{D}'Ascoli}
  et~al.}{2018}]{minidiscs_2}
{\lowercase{D}'Ascoli} S.,  {Noble} S.~C.,  {Bowen} D.~B.,  {Campanelli} M.,
  {Krolik} J.~H.,   {Mewes} V.,  2018, \mn@doi [\apj]
  {10.3847/1538-4357/aad8b4}, 865, 140

\bibitem[\protect\citeauthoryear{Davies \& Lin}{Davies \&
  Lin}{2020}]{Davies_2020}
Davies M.~B.,  Lin D. N.~C.,  2020, \mn@doi [Monthly Notices of the Royal
  Astronomical Society] {10.1093/mnras/staa2590}, 498, 3452–3456

\bibitem[\protect\citeauthoryear{{Davis} \& {El-Abd}}{{Davis} \&
  {El-Abd}}{2019}]{2019ApJ...874...23D}
{Davis} S.~W.,  {El-Abd} S.,  2019, \mn@doi [\apj] {10.3847/1538-4357/ab05c5},
  \href {https://ui.adsabs.harvard.edu/abs/2019ApJ...874...23D} {874, 23}

\bibitem[\protect\citeauthoryear{{Derdzinski}, {D'Orazio}, {Duffell}, {Haiman}
  \& {MacFadyen}}{{Derdzinski} et~al.}{2021}]{2021MNRAS.501.3540D}
{Derdzinski} A.,  {D'Orazio} D.,  {Duffell} P.,  {Haiman} Z.,   {MacFadyen} A.,
   2021, \mn@doi [\mnras] {10.1093/mnras/staa3976}, \href
  {https://ui.adsabs.harvard.edu/abs/2021MNRAS.501.3540D} {501, 3540}

\bibitem[\protect\citeauthoryear{{Dewangan}, {Tripathi}, {Papadakis}  \&
  {Singh}}{{Dewangan} et~al.}{2021}]{2021MNRAS.504.4015D}
{Dewangan} G.~C.,  {Tripathi} P.,  {Papadakis} I.~E.,   {Singh} K.~P.,  2021,
  \mn@doi [\mnras] {10.1093/mnras/stab1113}, \href
  {https://ui.adsabs.harvard.edu/abs/2021MNRAS.504.4015D} {504, 4015}

\bibitem[\protect\citeauthoryear{{Diamond-Stanic}, {Rieke}  \&
  {Rigby}}{{Diamond-Stanic} et~al.}{2009}]{diamond2009}
{Diamond-Stanic} A.~M.,  {Rieke} G.~H.,   {Rigby} J.~R.,  2009, \mn@doi [\apj]
  {10.1088/0004-637X/698/1/623}, \href
  {https://ui.adsabs.harvard.edu/abs/2009ApJ...698..623D} {698, 623}

\bibitem[\protect\citeauthoryear{{Done}, {Davis}, {Jin}, {Blaes}  \&
  {Ward}}{{Done} et~al.}{2012}]{2012MNRAS.420.1848D}
{Done} C.,  {Davis} S.~W.,  {Jin} C.,  {Blaes} O.,   {Ward} M.,  2012, \mn@doi
  [\mnras] {10.1111/j.1365-2966.2011.19779.x}, \href
  {https://ui.adsabs.harvard.edu/abs/2012MNRAS.420.1848D} {420, 1848}

\bibitem[\protect\citeauthoryear{{Eardley} \& {Lightman}}{{Eardley} \&
  {Lightman}}{1975}]{1975ApJ...200..187E}
{Eardley} D.~M.,  {Lightman} A.~P.,  1975, \mn@doi [\apj] {10.1086/153777},
  \href {https://ui.adsabs.harvard.edu/abs/1975ApJ...200..187E} {200, 187}

\bibitem[\protect\citeauthoryear{{Erickcek}, {Kamionkowski}  \&
  {Benson}}{{Erickcek} et~al.}{2006}]{2006MNRAS.371.1992E}
{Erickcek} A.~L.,  {Kamionkowski} M.,   {Benson} A.~J.,  2006, \mn@doi [\mnras]
  {10.1111/j.1365-2966.2006.10838.x}, \href
  {https://ui.adsabs.harvard.edu/abs/2006MNRAS.371.1992E} {371, 1992}

\bibitem[\protect\citeauthoryear{{Farris}, {Duffell}, {MacFadyen}  \&
  {Haiman}}{{Farris} et~al.}{2015}]{minidiscs_1}
{Farris} B.~D.,  {Duffell} P.,  {MacFadyen} A.~I.,   {Haiman} Z.,  2015,
  \mn@doi [\mnras] {10.1093/mnrasl/slu160}, \href
  {https://ui.adsabs.harvard.edu/abs/2015MNRAS.446L..36F} {446, L36}

\bibitem[\protect\citeauthoryear{{Fawcett}, {Alexander}, {Rosario}, {Klindt},
  {Lusso}, {Morabito}  \& {Calistro Rivera}}{{Fawcett}
  et~al.}{2022}]{fawcet2022}
{Fawcett} V.~A.,  {Alexander} D.~M.,  {Rosario} D.~J.,  {Klindt} L.,  {Lusso}
  E.,  {Morabito} L.~K.,   {Calistro Rivera} G.,  2022, \mn@doi [\mnras]
  {10.1093/mnras/stac945}, \href
  {https://ui.adsabs.harvard.edu/abs/2022MNRAS.513.1254F} {513, 1254}

\bibitem[\protect\citeauthoryear{{Ferreira}, {Petrucci}, {Henri}, {Saug{\'e}}
  \& {Pelletier}}{{Ferreira} et~al.}{2006}]{ferreira2006}
{Ferreira} J.,  {Petrucci} P.~O.,  {Henri} G.,  {Saug{\'e}} L.,   {Pelletier}
  G.,  2006, \mn@doi [\aap] {10.1051/0004-6361:20052689}, \href
  {https://ui.adsabs.harvard.edu/abs/2006A&A...447..813F} {447, 813}

\bibitem[\protect\citeauthoryear{{Fischer}, {Crenshaw}, {Kraemer}  \&
  {Schmitt}}{{Fischer} et~al.}{2013}]{fischer2013}
{Fischer} T.~C.,  {Crenshaw} D.~M.,  {Kraemer} S.~B.,   {Schmitt} H.~R.,  2013,
  \mn@doi [\apjs] {10.1088/0067-0049/209/1/1}, \href
  {https://ui.adsabs.harvard.edu/abs/2013ApJS..209....1F} {209, 1}

\bibitem[\protect\citeauthoryear{{Fishbone} \& {Moncrief}}{{Fishbone} \&
  {Moncrief}}{1976}]{fishborne76}
{Fishbone} L.~G.,  {Moncrief} V.,  1976, \mn@doi [\apj] {10.1086/154565}, \href
  {https://ui.adsabs.harvard.edu/abs/1976ApJ...207..962F} {207, 962}

\bibitem[\protect\citeauthoryear{{Frank}, {King}  \& {Raine}}{{Frank}
  et~al.}{2002}]{Frank}
{Frank} J.,  {King} A.,   {Raine} D.,  2002, {Accretion Power in Astrophysics}.
Cambridge University Press

\bibitem[\protect\citeauthoryear{{Gaskell}, {Goosmann}, {Antonucci}  \&
  {Whysong}}{{Gaskell} et~al.}{2004}]{gaskell2004}
{Gaskell} C.~M.,  {Goosmann} R.~W.,  {Antonucci} R. R.~J.,   {Whysong} D.~H.,
  2004, \mn@doi [\apj] {10.1086/423885}, \href
  {https://ui.adsabs.harvard.edu/abs/2004ApJ...616..147G} {616, 147}

\bibitem[\protect\citeauthoryear{Gold, Paschalidis, Etienne, Shapiro  \&
  Pfeiffer}{Gold et~al.}{2014}]{Gold_2014}
Gold R.,  Paschalidis V.,  Etienne Z.~B.,  Shapiro S.~L.,   Pfeiffer H.~P.,
  2014, \mn@doi [Physical Review D] {10.1103/physrevd.89.064060}, 89

\bibitem[\protect\citeauthoryear{{Gordon}, {Clayton}, {Misselt}, {Landolt}  \&
  {Wolff}}{{Gordon} et~al.}{2003}]{gordon2003}
{Gordon} K.~D.,  {Clayton} G.~C.,  {Misselt} K.~A.,  {Landolt} A.~U.,   {Wolff}
  M.~J.,  2003, \mn@doi [\apj] {10.1086/376774}, \href
  {https://ui.adsabs.harvard.edu/abs/2003ApJ...594..279G} {594, 279}

\bibitem[\protect\citeauthoryear{{G{\"u}ltekin} \& {Miller}}{{G{\"u}ltekin} \&
  {Miller}}{2012}]{gultekin}
{G{\"u}ltekin} K.,  {Miller} J.~M.,  2012, \mn@doi [\apj]
  {10.1088/0004-637X/761/2/90}, 761, 90

\bibitem[\protect\citeauthoryear{{G{\"u}ltekin} et~al.,}{{G{\"u}ltekin}
  et~al.}{2009}]{2009ApJ...698..198G}
{G{\"u}ltekin} K.,  et~al., 2009, \mn@doi [\apj] {10.1088/0004-637X/698/1/198},
  \href {https://ui.adsabs.harvard.edu/abs/2009ApJ...698..198G} {698, 198}

\bibitem[\protect\citeauthoryear{Guo, Liu, Zafar  \& Liao}{Guo
  et~al.}{2019}]{Guo_2019}
Guo H.,  Liu X.,  Zafar T.,   Liao W.-T.,  2019, \mn@doi [\mnras]
  {10.1093/mnras/stz3566}, 492, 2910

\bibitem[\protect\citeauthoryear{{Hawking} \& {Israel}}{{Hawking} \&
  {Israel}}{1987}]{1987thyg.book.....H}
{Hawking} S.~W.,  {Israel} W.,  1987, {Three Hundred Years of Gravitation}.
Cambridge University Press

\bibitem[\protect\citeauthoryear{{Hills}}{{Hills}}{1975}]{Hills}
{Hills} J.~G.,  1975, \mn@doi [\nat] {10.1038/254295a0}, 254, 295

\bibitem[\protect\citeauthoryear{{Honma}}{{Honma}}{1996}]{1996PASJ...48...77H}
{Honma} F.,  1996, \mn@doi [\pasj] {10.1093/pasj/48.1.77}, \href
  {https://ui.adsabs.harvard.edu/abs/1996PASJ...48...77H} {48, 77}

\bibitem[\protect\citeauthoryear{{Hopkins} et~al.,}{{Hopkins}
  et~al.}{2004}]{hopkins2004}
{Hopkins} P.~F.,  et~al., 2004, \mn@doi [\aj] {10.1086/423291}, \href
  {https://ui.adsabs.harvard.edu/abs/2004AJ....128.1112H} {128, 1112}

\bibitem[\protect\citeauthoryear{{Huber}, {Suyu}, {Noebauer}, {Chan}, {Kromer},
  {Sim}, {Sluse}  \& {Taubenberger}}{{Huber} et~al.}{2021}]{huber2021}
{Huber} S.,  {Suyu} S.~H.,  {Noebauer} U.~M.,  {Chan} J.~H.~H.,  {Kromer} M.,
  {Sim} S.~A.,  {Sluse} D.,   {Taubenberger} S.,  2021, \mn@doi [\aap]
  {10.1051/0004-6361/202039218}, \href
  {https://ui.adsabs.harvard.edu/abs/2021A&A...646A.110H} {646, A110}

\bibitem[\protect\citeauthoryear{{Ichimaru}}{{Ichimaru}}{1977}]{ichimaru1977}
{Ichimaru} S.,  1977, \mn@doi [\apj] {10.1086/155314}, \href
  {https://ui.adsabs.harvard.edu/abs/1977ApJ...214..840I} {214, 840}

\bibitem[\protect\citeauthoryear{{Ivezi{\'c}} et~al.,}{{Ivezi{\'c}}
  et~al.}{2019}]{2019ApJ...873..111I}
{Ivezi{\'c}} {\v{Z}}.,  et~al., 2019, \mn@doi [\apj]
  {10.3847/1538-4357/ab042c}, \href
  {https://ui.adsabs.harvard.edu/abs/2019ApJ...873..111I} {873, 111}

\bibitem[\protect\citeauthoryear{{Kara} et~al.,}{{Kara}
  et~al.}{2021}]{kara2021agn}
{Kara} E.,  et~al., 2021, \mn@doi [\apj] {10.3847/1538-4357/ac2159}, \href
  {https://ui.adsabs.harvard.edu/abs/2021ApJ...922..151K} {922, 151}

\bibitem[\protect\citeauthoryear{{Karas} \& {{\v{S}}ubr}}{{Karas} \&
  {{\v{S}}ubr}}{2001}]{karas}
{Karas} V.,  {{\v{S}}ubr} L.,  2001, \mn@doi [\aap]
  {10.1051/0004-6361:20011009}, 376, 686

\bibitem[\protect\citeauthoryear{{Karas}, {Svoboda}  \&
  {Zaja{\v{c}}ek}}{{Karas} et~al.}{2021}]{2021bhns.confE...1K}
{Karas} V.,  {Svoboda} J.,   {Zaja{\v{c}}ek} M.,  2021, {Selected Chapters on
  Active Galactic Nuclei as Relativistic Systems}.
Silesian University, Opava (\mn@eprint {arXiv} {1901.06507})

\bibitem[\protect\citeauthoryear{{Kato} \& {Nakamura}}{{Kato} \&
  {Nakamura}}{1998}]{1998PASJ...50..559K}
{Kato} S.,  {Nakamura} K.~E.,  1998, \mn@doi [\pasj] {10.1093/pasj/50.6.559},
  \href {https://ui.adsabs.harvard.edu/abs/1998PASJ...50..559K} {50, 559}

\bibitem[\protect\citeauthoryear{{Kato}, {Fukue}  \& {Mineshige}}{{Kato}
  et~al.}{2008}]{2008bhad.book.....K}
{Kato} S.,  {Fukue} J.,   {Mineshige} S.,  2008, {Black-Hole Accretion Disks
  --- Towards a New Paradigm ---}.
Kyoto University Press: Kyoto

\bibitem[\protect\citeauthoryear{{Kelley}, {D'Orazio}  \& {Di
  Stefano}}{{Kelley} et~al.}{2021}]{2021MNRAS.508.2524K}
{Kelley} L.~Z.,  {D'Orazio} D.~J.,   {Di Stefano} R.,  2021, \mn@doi [\mnras]
  {10.1093/mnras/stab2776}, \href
  {https://ui.adsabs.harvard.edu/abs/2021MNRAS.508.2524K} {508, 2524}

\bibitem[\protect\citeauthoryear{{King} \& {Done}}{{King} \&
  {Done}}{1993}]{1993MNRAS.264..388K}
{King} A.~R.,  {Done} C.,  1993, \mn@doi [\mnras] {10.1093/mnras/264.2.388},
  \href {https://ui.adsabs.harvard.edu/abs/1993MNRAS.264..388K} {264, 388}

\bibitem[\protect\citeauthoryear{{Kormendy} \& {Ho}}{{Kormendy} \&
  {Ho}}{2013}]{2013ARA&A..51..511K}
{Kormendy} J.,  {Ho} L.~C.,  2013, \mn@doi [\araa]
  {10.1146/annurev-astro-082708-101811}, \href
  {https://ui.adsabs.harvard.edu/abs/2013ARA&A..51..511K} {51, 511}

\bibitem[\protect\citeauthoryear{Krolik, Hawley  \& Hirose}{Krolik
  et~al.}{2005}]{Krolik_2005}
Krolik J.~H.,  Hawley J.~F.,   Hirose S.,  2005, \mn@doi [\apj]
  {10.1086/427932}, 622, 1008

\bibitem[\protect\citeauthoryear{{Kumar}, {Dewangan}, {Singh}, {Gandhi},
  {Papadakis}, {Tripathi}  \& {Mallick}}{{Kumar}
  et~al.}{2023}]{2023arXiv230311882K}
{Kumar} S.,  {Dewangan} G.~C.,  {Singh} K.~P.,  {Gandhi} P.,  {Papadakis}
  I.~E.,  {Tripathi} P.,   {Mallick} L.,  2023, \mn@doi [arXiv e-prints]
  {10.48550/arXiv.2303.11882}, \href
  {https://ui.adsabs.harvard.edu/abs/2023arXiv230311882K} {p. arXiv:2303.11882}

\bibitem[\protect\citeauthoryear{{LSST Science Collaboration} et~al.,}{{LSST
  Science Collaboration} et~al.}{2009}]{LSST_SB2009}
{LSST Science Collaboration} et~al., 2009, arXiv e-prints, \href
  {https://ui.adsabs.harvard.edu/abs/2009arXiv0912.0201L} {p. arXiv:0912.0201}

\bibitem[\protect\citeauthoryear{{Lan{\v{c}}ov{\'a}}
  et~al.,}{{Lan{\v{c}}ov{\'a}} et~al.}{2019}]{2019ApJ...884L..37L}
{Lan{\v{c}}ov{\'a}} D.,  et~al., 2019, \mn@doi [\apjl]
  {10.3847/2041-8213/ab48f5}, \href
  {https://ui.adsabs.harvard.edu/abs/2019ApJ...884L..37L} {884, L37}

\bibitem[\protect\citeauthoryear{{Lawrence}}{{Lawrence}}{2012}]{lawrence2012}
{Lawrence} A.,  2012, \mn@doi [\mnras] {10.1111/j.1365-2966.2012.20889.x},
  \href {https://ui.adsabs.harvard.edu/abs/2012MNRAS.423..451L} {423, 451}

\bibitem[\protect\citeauthoryear{{Lawrence} \& {Elvis}}{{Lawrence} \&
  {Elvis}}{2010}]{lawrence2010}
{Lawrence} A.,  {Elvis} M.,  2010, \mn@doi [\apj]
  {10.1088/0004-637X/714/1/561}, \href
  {https://ui.adsabs.harvard.edu/abs/2010ApJ...714..561L} {714, 561}

\bibitem[\protect\citeauthoryear{{Liang}, {Shu}, {Wang}, {Tan}, {Zhang}, {Sun},
  {Jiang}  \& {Dou}}{{Liang} et~al.}{2022}]{ngc1566_gap_2022}
{Liang} W.~C.,  {Shu} X.~W.,  {Wang} J.~X.,  {Tan} Y.,  {Zhang} W.~J.,  {Sun}
  L.~M.,  {Jiang} N.,   {Dou} L.~M.,  2022, \mn@doi [J. of High Energy
  Astrophys.] {10.1016/j.jheap.2022.01.002}, \href
  {https://ui.adsabs.harvard.edu/abs/2022JHEAp..33...20L} {33, 20}

\bibitem[\protect\citeauthoryear{{Lin} \& {Papaloizou}}{{Lin} \&
  {Papaloizou}}{1986}]{1986ApJ...309..846L}
{Lin} D.~N.~C.,  {Papaloizou} J.,  1986, \mn@doi [\apj] {10.1086/164653}, \href
  {https://ui.adsabs.harvard.edu/abs/1986ApJ...309..846L} {309, 846}

\bibitem[\protect\citeauthoryear{{Liu}, {Meyer}  \& {Meyer-Hofmeister}}{{Liu}
  et~al.}{2006}]{below_adaf}
{Liu} B.~F.,  {Meyer} F.,   {Meyer-Hofmeister} E.,  2006, \mn@doi [\aap]
  {10.1051/0004-6361:20065430}, \href
  {https://ui.adsabs.harvard.edu/abs/2006A&A...454L...9L} {454, L9}

\bibitem[\protect\citeauthoryear{{Liu} et~al.,}{{Liu} et~al.}{2019}]{liu2019}
{Liu} T.,  et~al., 2019, \mn@doi [\apj] {10.3847/1538-4357/ab40cb}, \href
  {https://ui.adsabs.harvard.edu/abs/2019ApJ...884...36L} {884, 36}

\bibitem[\protect\citeauthoryear{{Lusso} \& {Risaliti}}{{Lusso} \&
  {Risaliti}}{2016}]{lusso2016}
{Lusso} E.,  {Risaliti} G.,  2016, \mn@doi [\apj]
  {10.3847/0004-637X/819/2/154}, \href
  {https://ui.adsabs.harvard.edu/abs/2016ApJ...819..154L} {819, 154}

\bibitem[\protect\citeauthoryear{{Lusso} et~al.,}{{Lusso}
  et~al.}{2020}]{2020A&A...642A.150L}
{Lusso} E.,  et~al., 2020, \mn@doi [\aap] {10.1051/0004-6361/202038899}, \href
  {https://ui.adsabs.harvard.edu/abs/2020A&A...642A.150L} {642, A150}

\bibitem[\protect\citeauthoryear{{Lynden-Bell} \& {Pringle}}{{Lynden-Bell} \&
  {Pringle}}{1974}]{1974MNRAS.168..603L}
{Lynden-Bell} D.,  {Pringle} J.~E.,  1974, \mn@doi [\mnras]
  {10.1093/mnras/168.3.603}, \href
  {https://ui.adsabs.harvard.edu/abs/1974MNRAS.168..603L} {168, 603}

\bibitem[\protect\citeauthoryear{{Magorrian} et~al.,}{{Magorrian}
  et~al.}{1998}]{magorian}
{Magorrian} J.,  et~al., 1998, \mn@doi [\aj] {10.1086/300353}, 115, 2285

\bibitem[\protect\citeauthoryear{{Marcel} et~al.,}{{Marcel}
  et~al.}{2019}]{marcel2019}
{Marcel} G.,  et~al., 2019, \mn@doi [\aap] {10.1051/0004-6361/201935060}, \href
  {https://ui.adsabs.harvard.edu/abs/2019A&A...626A.115M} {626, A115}

\bibitem[\protect\citeauthoryear{{Marculewicz}, {Nikolajuk}  \&
  {R{\'o}{\.z}a{\'n}ska}}{{Marculewicz} et~al.}{2022}]{marculewicz2022}
{Marculewicz} M.,  {Nikolajuk} M.,   {R{\'o}{\.z}a{\'n}ska} A.,  2022, arXiv
  e-prints, \href {https://ui.adsabs.harvard.edu/abs/2022arXiv221010438M} {p.
  arXiv:2210.10438}

\bibitem[\protect\citeauthoryear{{Marin}}{{Marin}}{2014}]{marin2014}
{Marin} F.,  2014, \mn@doi [\mnras] {10.1093/mnras/stu593}, \href
  {https://ui.adsabs.harvard.edu/abs/2014MNRAS.441..551M} {441, 551}

\bibitem[\protect\citeauthoryear{{Mej{\'\i}a-Restrepo}, {Lira}, {Netzer},
  {Trakhtenbrot}  \& {Capellupo}}{{Mej{\'\i}a-Restrepo}
  et~al.}{2018}]{mejia2018}
{Mej{\'\i}a-Restrepo} J.~E.,  {Lira} P.,  {Netzer} H.,  {Trakhtenbrot} B.,
  {Capellupo} D.~M.,  2018, \mn@doi [Nature Astronomy]
  {10.1038/s41550-017-0305-z}, \href
  {https://ui.adsabs.harvard.edu/abs/2018NatAs...2...63M} {2, 63}

\bibitem[\protect\citeauthoryear{{Merritt}}{{Merritt}}{2013}]{2013degn.book.....M}
{Merritt} D.,  2013, {Dynamics and Evolution of Galactic Nuclei (Princeton:
  Princeton University Press)}.
Princeton University Press

\bibitem[\protect\citeauthoryear{{Metzger}, {Stone}  \& {Gilbaum}}{{Metzger}
  et~al.}{2022}]{2022ApJ...926..101M}
{Metzger} B.~D.,  {Stone} N.~C.,   {Gilbaum} S.,  2022, \mn@doi [\apj]
  {10.3847/1538-4357/ac3ee1}, \href
  {https://ui.adsabs.harvard.edu/abs/2022ApJ...926..101M} {926, 101}

\bibitem[\protect\citeauthoryear{{Meyer} \& {Meyer-Hofmeister}}{{Meyer} \&
  {Meyer-Hofmeister}}{2002}]{2002A&A...392L...5M}
{Meyer} F.,  {Meyer-Hofmeister} E.,  2002, \mn@doi [\aap]
  {10.1051/0004-6361:20021075}, \href
  {https://ui.adsabs.harvard.edu/abs/2002A&A...392L...5M} {392, L5}

\bibitem[\protect\citeauthoryear{{Miranda} \& {Rafikov}}{{Miranda} \&
  {Rafikov}}{2020}]{2020ApJ...904..121M}
{Miranda} R.,  {Rafikov} R.~R.,  2020, \mn@doi [\apj]
  {10.3847/1538-4357/abbee7}, \href
  {https://ui.adsabs.harvard.edu/abs/2020ApJ...904..121M} {904, 121}

\bibitem[\protect\citeauthoryear{{Misner}, {Thorne}  \& {Wheeler}}{{Misner}
  et~al.}{1973}]{misner}
{Misner} C.~W.,  {Thorne} K.~S.,   {Wheeler} J.~A.,  1973, {Gravitation}.
San Francisco: W.H. Freeman and Co

\bibitem[\protect\citeauthoryear{{Moranchel-Basurto}, {S{\'a}nchez-Salcedo},
  {Chametla}  \& {Vel{\'a}zquez}}{{Moranchel-Basurto}
  et~al.}{2021}]{2021ApJ...906...15M}
{Moranchel-Basurto} A.,  {S{\'a}nchez-Salcedo} F.~J.,  {Chametla} R.~O.,
  {Vel{\'a}zquez} P.~F.,  2021, \mn@doi [\apj] {10.3847/1538-4357/abca88},
  \href {https://ui.adsabs.harvard.edu/abs/2021ApJ...906...15M} {906, 15}

\bibitem[\protect\citeauthoryear{{Morrissey} et~al.,}{{Morrissey}
  et~al.}{2007}]{morrissey2007}
{Morrissey} P.,  et~al., 2007, \mn@doi [\apjs] {10.1086/520512}, \href
  {https://ui.adsabs.harvard.edu/abs/2007ApJS..173..682M} {173, 682}

\bibitem[\protect\citeauthoryear{{Narayan}}{{Narayan}}{2000}]{2000ApJ...536..663N}
{Narayan} R.,  2000, \mn@doi [\apj] {10.1086/308956}, \href
  {https://ui.adsabs.harvard.edu/abs/2000ApJ...536..663N} {536, 663}

\bibitem[\protect\citeauthoryear{{Narayan} \& {Yi}}{{Narayan} \&
  {Yi}}{1994}]{1994ApJ...428L..13N}
{Narayan} R.,  {Yi} I.,  1994, \mn@doi [\apjl] {10.1086/187381}, \href
  {https://ui.adsabs.harvard.edu/abs/1994ApJ...428L..13N} {428, L13}

\bibitem[\protect\citeauthoryear{{Narayan}, {Igumenshchev}  \&
  {Abramowicz}}{{Narayan} et~al.}{2003}]{2003PASJ...55L..69N}
{Narayan} R.,  {Igumenshchev} I.~V.,   {Abramowicz} M.~A.,  2003, \mn@doi
  [\pasj] {10.1093/pasj/55.6.L69}, \href
  {https://ui.adsabs.harvard.edu/abs/2003PASJ...55L..69N} {55, L69}

\bibitem[\protect\citeauthoryear{{Netzer}}{{Netzer}}{2019}]{netzer}
{Netzer} H.,  2019, \mn@doi [\mnras] {10.1093/mnras/stz2016}, \href
  {https://ui.adsabs.harvard.edu/abs/2019MNRAS.488.5185N} {488, 5185}

\bibitem[\protect\citeauthoryear{{O'Leary}, {Moster}, {Naab}  \&
  {Somerville}}{{O'Leary} et~al.}{2021}]{2021MNRAS.501.3215O}
{O'Leary} J.~A.,  {Moster} B.~P.,  {Naab} T.,   {Somerville} R.~S.,  2021,
  \mn@doi [\mnras] {10.1093/mnras/staa3746}, \href
  {https://ui.adsabs.harvard.edu/abs/2021MNRAS.501.3215O} {501, 3215}

\bibitem[\protect\citeauthoryear{{Page} \& {Thorne}}{{Page} \&
  {Thorne}}{1974}]{1974ApJ...191..499P}
{Page} D.~N.,  {Thorne} K.~S.,  1974, \mn@doi [\apj] {10.1086/152990}, \href
  {https://ui.adsabs.harvard.edu/abs/1974ApJ...191..499P} {191, 499}

\bibitem[\protect\citeauthoryear{{Peng} \& {Chen}}{{Peng} \&
  {Chen}}{2021}]{2021MNRAS.505.1324P}
{Peng} P.,  {Chen} X.,  2021, \mn@doi [\mnras] {10.1093/mnras/stab1419}, \href
  {https://ui.adsabs.harvard.edu/abs/2021MNRAS.505.1324P} {505, 1324}

\bibitem[\protect\citeauthoryear{{Peters}}{{Peters}}{1964}]{1964PhRv..136.1224P}
{Peters} P.~C.,  1964, \mn@doi [Physical Review] {10.1103/PhysRev.136.B1224},
  \href {https://ui.adsabs.harvard.edu/abs/1964PhRv..136.1224P} {136, 1224}

\bibitem[\protect\citeauthoryear{{Poole} et~al.,}{{Poole}
  et~al.}{2008}]{poole2008}
{Poole} T.~S.,  et~al., 2008, \mn@doi [\mnras]
  {10.1111/j.1365-2966.2007.12563.x}, \href
  {https://ui.adsabs.harvard.edu/abs/2008MNRAS.383..627P} {383, 627}

\bibitem[\protect\citeauthoryear{{Prince}, {Hryniewicz}, {Panda}, {Czerny}  \&
  {Pollo}}{{Prince} et~al.}{2022}]{prince2022}
{Prince} R.,  {Hryniewicz} K.,  {Panda} S.,  {Czerny} B.,   {Pollo} A.,  2022,
  \mn@doi [\apj] {10.3847/1538-4357/ac3f36}, \href
  {https://ui.adsabs.harvard.edu/abs/2022ApJ...925..215P} {925, 215}

\bibitem[\protect\citeauthoryear{{Pugliese} \& {Stuchl{\'\i}k}}{{Pugliese} \&
  {Stuchl{\'\i}k}}{2015}]{2015ApJS..221...25P}
{Pugliese} D.,  {Stuchl{\'\i}k} Z.,  2015, \mn@doi [\apjs]
  {10.1088/0067-0049/221/2/25}, \href
  {https://ui.adsabs.harvard.edu/abs/2015ApJS..221...25P} {221, 25}

\bibitem[\protect\citeauthoryear{{Ramos Padilla}, {Wang}, {Ma{\l}ek},
  {Efstathiou}  \& {Yang}}{{Ramos Padilla} et~al.}{2022}]{ramos2022}
{Ramos Padilla} A.~F.,  {Wang} L.,  {Ma{\l}ek} K.,  {Efstathiou} A.,   {Yang}
  G.,  2022, \mn@doi [\mnras] {10.1093/mnras/stab3486}, \href
  {https://ui.adsabs.harvard.edu/abs/2022MNRAS.510..687R} {510, 687}

\bibitem[\protect\citeauthoryear{{Rees}}{{Rees}}{1988}]{Rees}
{Rees} M.~J.,  1988, \mn@doi [\nat] {10.1038/333523a0}, 333, 523

\bibitem[\protect\citeauthoryear{Reynolds}{Reynolds}{2003}]{Reynolds_2003}
Reynolds C.,  2003, \mn@doi [Physics Reports] {10.1016/s0370-1573(02)00584-7},
  377, 389

\bibitem[\protect\citeauthoryear{{Reynolds} \& {Begelman}}{{Reynolds} \&
  {Begelman}}{1997}]{1997ApJ...488..109R}
{Reynolds} C.~S.,  {Begelman} M.~C.,  1997, \mn@doi [\apj] {10.1086/304703},
  \href {https://ui.adsabs.harvard.edu/abs/1997ApJ...488..109R} {488, 109}

\bibitem[\protect\citeauthoryear{{Ricci} et~al.,}{{Ricci} et~al.}{2020}]{Ricci}
{Ricci} C.,  et~al., 2020, \mn@doi [\apjl] {10.3847/2041-8213/ab91a1}, 898, L1

\bibitem[\protect\citeauthoryear{{Riffert}}{{Riffert}}{2000}]{2000ApJ...529..119R}
{Riffert} H.,  2000, \mn@doi [\apj] {10.1086/308248}, \href
  {https://ui.adsabs.harvard.edu/abs/2000ApJ...529..119R} {529, 119}

\bibitem[\protect\citeauthoryear{{Risaliti} \& {Lusso}}{{Risaliti} \&
  {Lusso}}{2015}]{risaliti2015}
{Risaliti} G.,  {Lusso} E.,  2015, \mn@doi [\apj] {10.1088/0004-637X/815/1/33},
  \href {https://ui.adsabs.harvard.edu/abs/2015ApJ...815...33R} {815, 33}

\bibitem[\protect\citeauthoryear{{Romero}, {Vila}  \& {P{\'e}rez}}{{Romero}
  et~al.}{2016}]{2016A&A...588A.125R}
{Romero} G.~E.,  {Vila} G.~S.,   {P{\'e}rez} D.,  2016, \mn@doi [\aap]
  {10.1051/0004-6361/201527479}, \href
  {https://ui.adsabs.harvard.edu/abs/2016A&A...588A.125R} {588, A125}

\bibitem[\protect\citeauthoryear{{R\'o\.za\'nska} \& {Czerny}}{{R\'o\.za\'nska}
  \& {Czerny}}{2000}]{2000A&A...360.1170R}
{R\'o\.za\'nska} A.,  {Czerny} B.,  2000, \aap, \href
  {https://ui.adsabs.harvard.edu/abs/2000A&A...360.1170R} {360, 1170}

\bibitem[\protect\citeauthoryear{{Rusakov}, {Steinhardt}, {Schramm}, {Faisst},
  {Masters}, {Mobasher}  \& {Pattarakijwanich}}{{Rusakov}
  et~al.}{2023}]{2023ApJ...944..217R}
{Rusakov} V.,  {Steinhardt} C.~L.,  {Schramm} M.,  {Faisst} A.~L.,  {Masters}
  D.,  {Mobasher} B.,   {Pattarakijwanich} P.,  2023, \mn@doi [\apj]
  {10.3847/1538-4357/acadd8}, \href
  {https://ui.adsabs.harvard.edu/abs/2023ApJ...944..217R} {944, 217}

\bibitem[\protect\citeauthoryear{{Saxton}, {Komossa}, {Auchettl}  \&
  {Jonker}}{{Saxton} et~al.}{2020}]{2020SSRv..216...85S}
{Saxton} R.,  {Komossa} S.,  {Auchettl} K.,   {Jonker} P.~G.,  2020, \mn@doi
  [\ssr] {10.1007/s11214-020-00708-4}, \href
  {https://ui.adsabs.harvard.edu/abs/2020SSRv..216...85S} {216, 85}

\bibitem[\protect\citeauthoryear{{Schneider}}{{Schneider}}{2015}]{2015eaci.book.....S}
{Schneider} P.,  2015, {Extragalactic Astronomy and Cosmology: An
  Introduction}.
Springer Berlin, Heidelberg, \mn@doi{10.1007/978-3-642-54083-7}

\bibitem[\protect\citeauthoryear{{Schwarz}}{{Schwarz}}{1978}]{1978AnSta...6..461S}
{Schwarz} G.,  1978, Annals of Statistics, \href
  {https://ui.adsabs.harvard.edu/abs/1978AnSta...6..461S} {6, 461}

\bibitem[\protect\citeauthoryear{{Seaton}}{{Seaton}}{1979}]{seaton1979}
{Seaton} M.~J.,  1979, \mn@doi [\mnras] {10.1093/mnras/187.1.73P}, \href
  {https://ui.adsabs.harvard.edu/abs/1979MNRAS.187P..73S} {187, 73}

\bibitem[\protect\citeauthoryear{{Shakura} \& {Sunyaev}}{{Shakura} \&
  {Sunyaev}}{1973}]{shakura}
{Shakura} N.~I.,  {Sunyaev} R.~A.,  1973, \aap, \href
  {https://ui.adsabs.harvard.edu/abs/1973A&A....24..337S} {500, 33}

\bibitem[\protect\citeauthoryear{Shapiro}{Shapiro}{2010}]{Shapiro_2010}
Shapiro S.~L.,  2010, \mn@doi [Physical Review D] {10.1103/physrevd.81.024019},
  81

\bibitem[\protect\citeauthoryear{{Shapiro} \& {Teukolsky}}{{Shapiro} \&
  {Teukolsky}}{1983}]{shapiro_and_teukolsky_1983}
{Shapiro} S.~L.,  {Teukolsky} S.~A.,  1983, {Black Holes, White Dwarfs, and
  Neutron Stars: the Physics of Compact Objects}.
New York: Wiley

\bibitem[\protect\citeauthoryear{{Shi} \& {Krolik}}{{Shi} \&
  {Krolik}}{2016}]{2016ApJ...832...22S}
{Shi} J.-M.,  {Krolik} J.~H.,  2016, \mn@doi [\apj]
  {10.3847/0004-637X/832/1/22}, \href
  {https://ui.adsabs.harvard.edu/abs/2016ApJ...832...22S} {832, 22}

\bibitem[\protect\citeauthoryear{{\'{S}niegowska}, {Czerny}, {Bon}  \&
  {Bon}}{{\'{S}niegowska} et~al.}{2020}]{sniegowska}
{\'{S}niegowska} M.,  {Czerny} B.,  {Bon} E.,   {Bon} N.,  2020, \mn@doi [\aap]
  {10.1051/0004-6361/202038575}, \href
  {https://ui.adsabs.harvard.edu/abs/2020A&A...641A.167S} {641, A167}

\bibitem[\protect\citeauthoryear{{{\'S}niegowska}, {Grz{\k{e}}dzielski},
  {Czerny}  \& {Janiuk}}{{{\'S}niegowska} et~al.}{2022}]{sniegowska22}
{{\'S}niegowska} M.,  {Grz{\k{e}}dzielski} M.,  {Czerny} B.,   {Janiuk} A.,
  2022, submitted to A\&A, \href
  {https://ui.adsabs.harvard.edu/abs/2022arXiv220410067S} {p. arXiv:2204.10067}

\bibitem[\protect\citeauthoryear{{Sochora}, {Karas}, {Svoboda}  \&
  {Dov{\v{c}}iak}}{{Sochora} et~al.}{2011}]{2011MNRAS.418..276S}
{Sochora} V.,  {Karas} V.,  {Svoboda} J.,   {Dov{\v{c}}iak} M.,  2011, \mn@doi
  [\mnras] {10.1111/j.1365-2966.2011.19483.x}, \href
  {https://ui.adsabs.harvard.edu/abs/2011MNRAS.418..276S} {418, 276}

\bibitem[\protect\citeauthoryear{{Soltan}}{{Soltan}}{1982}]{1982MNRAS.200..115S}
{Soltan} A.,  1982, \mn@doi [\mnras] {10.1093/mnras/200.1.115}, \href
  {https://ui.adsabs.harvard.edu/abs/1982MNRAS.200..115S} {200, 115}

\bibitem[\protect\citeauthoryear{{Stoeger}}{{Stoeger}}{1976}]{1976A&A....53..267S}
{Stoeger} W.~R.,  1976, \aap, \href
  {https://ui.adsabs.harvard.edu/abs/1976A&A....53..267S} {53, 267}

\bibitem[\protect\citeauthoryear{{\v{S}tolc} \& {Karas}}{{\v{S}tolc} \&
  {Karas}}{2019}]{2019AN....340..570S}
{\v{S}tolc} M.,  {Karas} V.,  2019, \mn@doi [Astronomische Nachrichten]
  {10.1002/asna.201913658}, \href
  {https://ui.adsabs.harvard.edu/abs/2019AN....340..570S} {340, 570}

\bibitem[\protect\citeauthoryear{{\v{S}tolc}, {Karas}, {Sukov\'a}, {Witzany}
  \& {Zaja\v{c}ek}}{{\v{S}tolc} et~al.}{2020}]{wds_stolc}
{\v{S}tolc} M.,  {Karas} V.,  {Sukov\'a} P.,  {Witzany} V.,   {Zaja\v{c}ek} M.,
   2020, in WDS’20 Proceedings of Contributed Papers — Physics (eds. J.
  Šafránková and J. Pavlů), pp. 59-64,Matfzypress,Prague

\bibitem[\protect\citeauthoryear{{\v{S}ubr} \& {Karas}}{{\v{S}ubr} \&
  {Karas}}{1999}]{subr}
{\v{S}ubr} L.,  {Karas} V.,  1999, \aap, 352, 452

\bibitem[\protect\citeauthoryear{{Sukov{\'a}}, {Zaja{\v{c}}ek}, {Witzany}  \&
  {Karas}}{{Sukov{\'a}} et~al.}{2020}]{2020arXiv201202608S}
{Sukov{\'a}} P.,  {Zaja{\v{c}}ek} M.,  {Witzany} V.,   {Karas} V.,  2020, in
  Proceedings of RAGtime 20-22 (Silesian University, Opava), p. 299, Silesian
  University in Opava,Opava

\bibitem[\protect\citeauthoryear{{Sukov{\'a}}, {Zaja{\v{c}}ek}, {Witzany}  \&
  {Karas}}{{Sukov{\'a}} et~al.}{2021}]{2021ApJ...917...43S}
{Sukov{\'a}} P.,  {Zaja{\v{c}}ek} M.,  {Witzany} V.,   {Karas} V.,  2021,
  \mn@doi [\apj] {10.3847/1538-4357/ac05c6}, \href
  {https://ui.adsabs.harvard.edu/abs/2021ApJ...917...43S} {917, 43}

\bibitem[\protect\citeauthoryear{{Syer}, {Clarke}  \& {Rees}}{{Syer}
  et~al.}{1991}]{1991MNRAS.250..505S}
{Syer} D.,  {Clarke} C.~J.,   {Rees} M.~J.,  1991, \mn@doi [\mnras]
  {10.1093/mnras/250.3.505}, \href
  {https://ui.adsabs.harvard.edu/abs/1991MNRAS.250..505S} {250, 505}

\bibitem[\protect\citeauthoryear{{Takeuchi}, {Miyama}  \& {Lin}}{{Takeuchi}
  et~al.}{1996}]{1996ApJ...460..832T}
{Takeuchi} T.,  {Miyama} S.~M.,   {Lin} D.~N.~C.,  1996, \mn@doi [\apj]
  {10.1086/177013}, \href
  {https://ui.adsabs.harvard.edu/abs/1996ApJ...460..832T} {460, 832}

\bibitem[\protect\citeauthoryear{{Tanaka} et~al.,}{{Tanaka}
  et~al.}{1995}]{tanaka95}
{Tanaka} Y.,  et~al., 1995, \mn@doi [\nat] {10.1038/375659a0}, \href
  {https://ui.adsabs.harvard.edu/abs/1995Natur.375..659T} {375, 659}

\bibitem[\protect\citeauthoryear{{Tananbaum} et~al.,}{{Tananbaum}
  et~al.}{1979}]{tananbaum1979}
{Tananbaum} H.,  et~al., 1979, \mn@doi [\apjl] {10.1086/183100}, \href
  {https://ui.adsabs.harvard.edu/abs/1979ApJ...234L...9T} {234, L9}

\bibitem[\protect\citeauthoryear{{Tursunov}, {Zaja{\v{c}}ek}, {Eckart},
  {Kolo{\v{s}}}, {Britzen}, {Stuchl{\'\i}k}, {Czerny}  \& {Karas}}{{Tursunov}
  et~al.}{2020}]{2020ApJ...897...99T}
{Tursunov} A.,  {Zaja{\v{c}}ek} M.,  {Eckart} A.,  {Kolo{\v{s}}} M.,  {Britzen}
  S.,  {Stuchl{\'\i}k} Z.,  {Czerny} B.,   {Karas} V.,  2020, \mn@doi [\apj]
  {10.3847/1538-4357/ab980e}, \href
  {https://ui.adsabs.harvard.edu/abs/2020ApJ...897...99T} {897, 99}

\bibitem[\protect\citeauthoryear{{Vestergaard} \& {Peterson}}{{Vestergaard} \&
  {Peterson}}{2006}]{Vestergaard2006}
{Vestergaard} M.,  {Peterson} B.~M.,  2006, \mn@doi [\apj] {10.1086/500572},
  \href {https://ui.adsabs.harvard.edu/abs/2006ApJ...641..689V} {641, 689}

\bibitem[\protect\citeauthoryear{{Wald}}{{Wald}}{1972}]{1972PhRvD...6.1476W}
{Wald} R.~M.,  1972, \mn@doi [\prd] {10.1103/PhysRevD.6.1476}, \href
  {https://ui.adsabs.harvard.edu/abs/1972PhRvD...6.1476W} {6, 1476}

\bibitem[\protect\citeauthoryear{{Ward}}{{Ward}}{1997}]{1997Icar..126..261W}
{Ward} W.~R.,  1997, \mn@doi [\icarus] {10.1006/icar.1996.5647}, \href
  {https://ui.adsabs.harvard.edu/abs/1997Icar..126..261W} {126, 261}

\bibitem[\protect\citeauthoryear{{Werner} et~al.,}{{Werner}
  et~al.}{2022}]{2022arXiv220705485W}
{Werner} N.,  et~al., 2022, in {den Herder} J.-W.~A.,  {Nikzad} S.,
  {Nakazawa} K.,  eds,  Society of Photo-Optical Instrumentation Engineers
  (SPIE) Conference Series Vol. 12181, Space Telescopes and Instrumentation
  2022: Ultraviolet to Gamma Ray. p. 121810B (\mn@eprint {arXiv} {2207.05485}),
  \mn@doi{10.1117/12.2629531}

\bibitem[\protect\citeauthoryear{{Wilkin}}{{Wilkin}}{1996}]{1996ApJ...459L..31W}
{Wilkin} F.~P.,  1996, \mn@doi [\apjl] {10.1086/309939}, \href
  {https://ui.adsabs.harvard.edu/abs/1996ApJ...459L..31W} {459, L31}

\bibitem[\protect\citeauthoryear{{Witt}, {Charisi}, {Taylor}  \&
  {Burke-Spolaor}}{{Witt} et~al.}{2021}]{witt2021}
{Witt} C.~A.,  {Charisi} M.,  {Taylor} S.~R.,   {Burke-Spolaor} S.,  2021,
  submitted to ApJ, \href
  {https://ui.adsabs.harvard.edu/abs/2021arXiv211007465W} {p. arXiv:2110.07465}

\bibitem[\protect\citeauthoryear{{Xin} \& {Haiman}}{{Xin} \&
  {Haiman}}{2021}]{2021MNRAS.506.2408X}
{Xin} C.,  {Haiman} Z.,  2021, \mn@doi [\mnras] {10.1093/mnras/stab1856}, \href
  {https://ui.adsabs.harvard.edu/abs/2021MNRAS.506.2408X} {506, 2408}

\bibitem[\protect\citeauthoryear{{Yuan} \& {Narayan}}{{Yuan} \&
  {Narayan}}{2014}]{yuan_naryan_2014}
{Yuan} F.,  {Narayan} R.,  2014, \mn@doi [\araa]
  {10.1146/annurev-astro-082812-141003}, \href
  {https://ui.adsabs.harvard.edu/abs/2014ARA&A..52..529Y} {52, 529}

\bibitem[\protect\citeauthoryear{{Zafar} et~al.,}{{Zafar}
  et~al.}{2015}]{zafar2015}
{Zafar} T.,  et~al., 2015, \mn@doi [\aap] {10.1051/0004-6361/201526570}, \href
  {https://ui.adsabs.harvard.edu/abs/2015A&A...584A.100Z} {584, A100}

\bibitem[\protect\citeauthoryear{{Zaja\v{c}ek}, {Karas}  \&
  {Kunneriath}}{{Zaja\v{c}ek} et~al.}{2015}]{2015AcPol..55..203Z}
{Zaja\v{c}ek} M.,  {Karas} V.,   {Kunneriath} D.,  2015, \mn@doi [Acta
  Polytechnica] {10.14311/AP.2015.55.0203}, \href
  {https://ui.adsabs.harvard.edu/abs/2015AcPol..55..203Z} {55, 203}

\bibitem[\protect\citeauthoryear{{Zaja{\v{c}}ek}, {Eckart}, {Karas},
  {Kunneriath}, {Shahzamanian}, {Sabha}, {Mu{\v{z}}i{\'c}}  \&
  {Valencia-S.}}{{Zaja{\v{c}}ek} et~al.}{2016}]{2016MNRAS.455.1257Z}
{Zaja{\v{c}}ek} M.,  {Eckart} A.,  {Karas} V.,  {Kunneriath} D.,
  {Shahzamanian} B.,  {Sabha} N.,  {Mu{\v{z}}i{\'c}} K.,   {Valencia-S.} M.,
  2016, \mn@doi [\mnras] {10.1093/mnras/stv2357}, \href
  {https://ui.adsabs.harvard.edu/abs/2016MNRAS.455.1257Z} {455, 1257}

\bibitem[\protect\citeauthoryear{{Zaja{\v{c}}ek}, {Tursunov}, {Eckart}  \&
  {Britzen}}{{Zaja{\v{c}}ek} et~al.}{2018}]{2018MNRAS.480.4408Z}
{Zaja{\v{c}}ek} M.,  {Tursunov} A.,  {Eckart} A.,   {Britzen} S.,  2018,
  \mn@doi [\mnras] {10.1093/mnras/sty2182}, \href
  {https://ui.adsabs.harvard.edu/abs/2018MNRAS.480.4408Z} {480, 4408}

\bibitem[\protect\citeauthoryear{{Zaja{\v{c}}ek} et~al.,}{{Zaja{\v{c}}ek}
  et~al.}{2021}]{2021ApJ...912...10Z}
{Zaja{\v{c}}ek} M.,  et~al., 2021, \mn@doi [\apj] {10.3847/1538-4357/abe9b2},
  \href {https://ui.adsabs.harvard.edu/abs/2021ApJ...912...10Z} {912, 10}

\bibitem[\protect\citeauthoryear{{Zaja{\v{c}}ek}, {Czerny}, {Sch{\"o}del},
  {Werner}  \& {Karas}}{{Zaja{\v{c}}ek} et~al.}{2022}]{2022NatAs...6.1008Z}
{Zaja{\v{c}}ek} M.,  {Czerny} B.,  {Sch{\"o}del} R.,  {Werner} N.,   {Karas}
  V.,  2022, \mn@doi [Nature Astronomy] {10.1038/s41550-022-01785-x}, \href
  {https://ui.adsabs.harvard.edu/abs/2022NatAs...6.1008Z} {6, 1008}

\bibitem[\protect\citeauthoryear{{Zdziarski} \& {De Marco}}{{Zdziarski} \& {De
  Marco}}{2020}]{2020ApJ...896L..36Z}
{Zdziarski} A.~A.,  {De Marco} B.,  2020, \mn@doi [\apjl]
  {10.3847/2041-8213/ab9899}, \href
  {https://ui.adsabs.harvard.edu/abs/2020ApJ...896L..36Z} {896, L36}

\bibitem[\protect\citeauthoryear{{Zdziarski}, {You}  \& {Szanecki}}{{Zdziarski}
  et~al.}{2022}]{2022ApJ...939L...2Z}
{Zdziarski} A.~A.,  {You} B.,   {Szanecki} M.,  2022, \mn@doi [\apjl]
  {10.3847/2041-8213/ac9474}, \href
  {https://ui.adsabs.harvard.edu/abs/2022ApJ...939L...2Z} {939, L2}

\bibitem[\protect\citeauthoryear{{{\v{S}}ubr} \& {Karas}}{{{\v{S}}ubr} \&
  {Karas}}{1999}]{1999A&A...352..452S}
{{\v{S}}ubr} L.,  {Karas} V.,  1999, \aap, \href
  {https://ui.adsabs.harvard.edu/abs/1999A&A...352..452S} {352, 452}

\bibitem[\protect\citeauthoryear{{{\v{S}}ubr}, {Karas}  \&
  {Hur{\'e}}}{{{\v{S}}ubr} et~al.}{2004}]{2004MNRAS.354.1177S}
{{\v{S}}ubr} L.,  {Karas} V.,   {Hur{\'e}} J.~M.,  2004, \mn@doi [\mnras]
  {10.1111/j.1365-2966.2004.08276.x}, \href
  {https://ui.adsabs.harvard.edu/abs/2004MNRAS.354.1177S} {354, 1177}

\makeatother
\end{thebibliography}




\appendix

\section{Stellar and pulsar stagnation radii}
\label{appendix_A}

\begin{figure*}
    \centering
    \includegraphics[width=0.9\columnwidth]{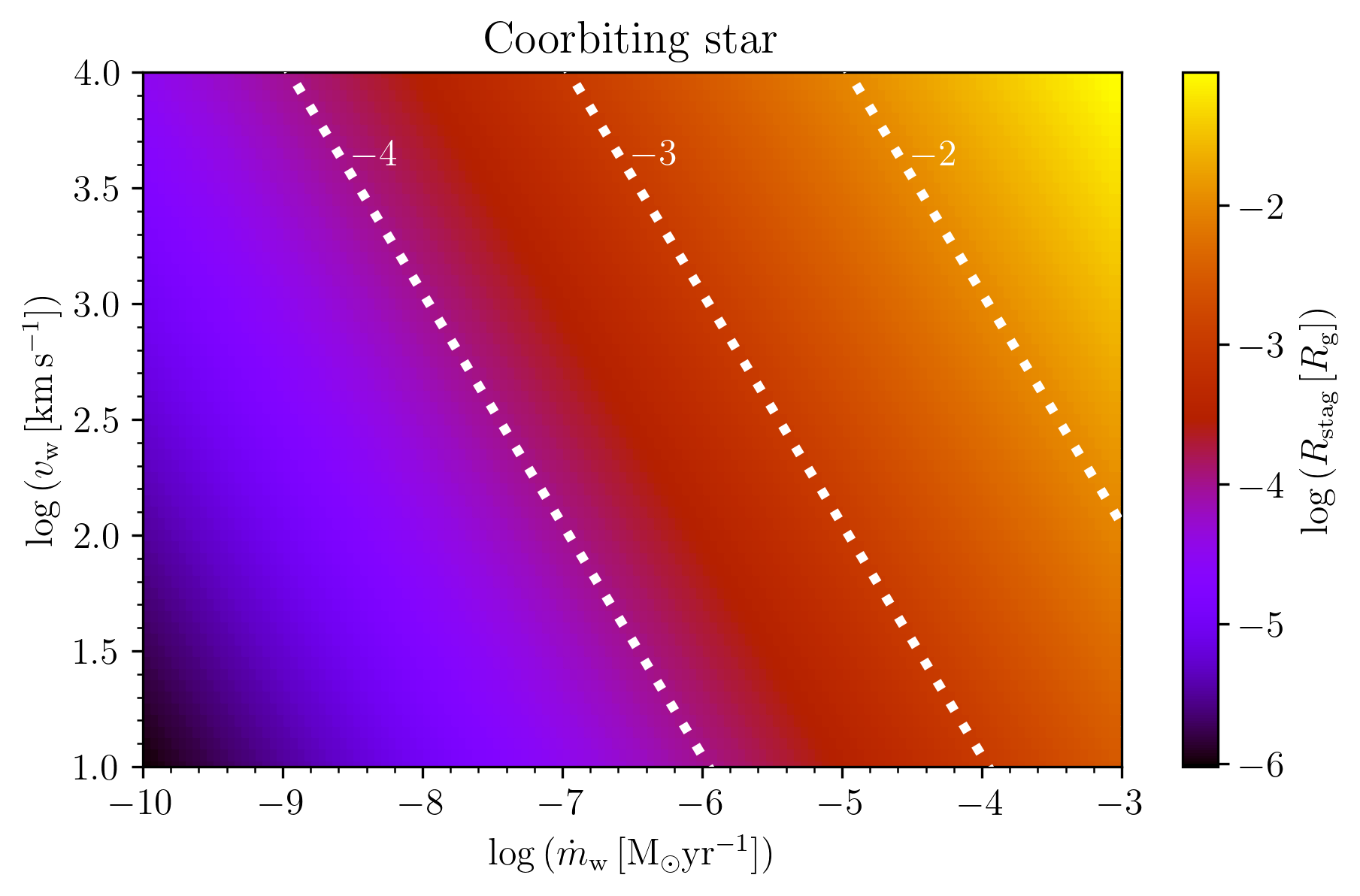} 
   \includegraphics[width=0.9\columnwidth]{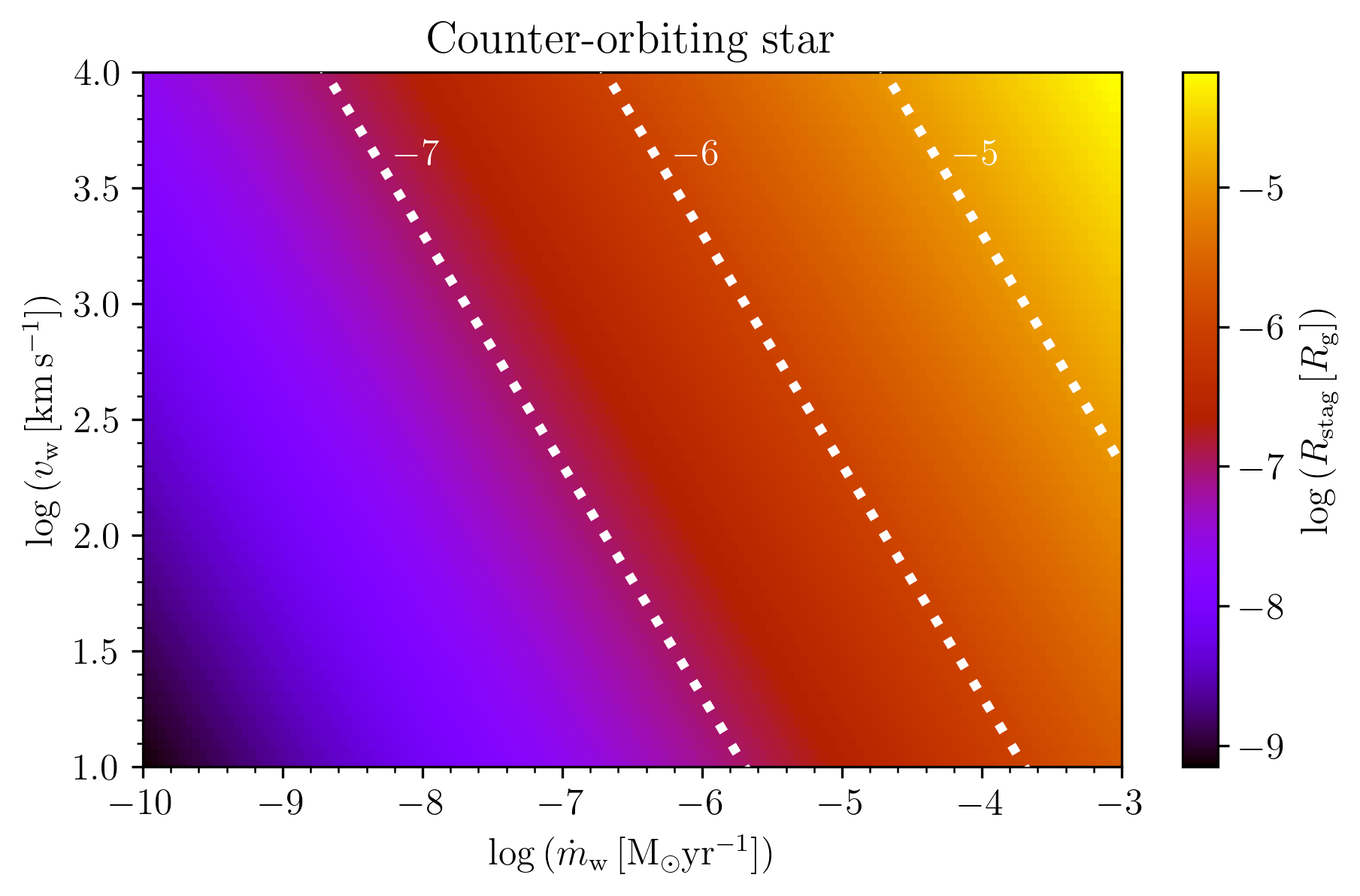}
   
    \caption{{ Stagnation radii of wind-blowing stars embedded in the accretion disc surrounding $10^9\,M_{\odot}$ SMBH with $\dot{m}=0.1$ and $\alpha=0.1$. The star orbits at the distance of $\sim 400\,R_{\rm g}$ from the SMBH. { Left panel:} Stagnation radius (expressed in gravitational radii) for the case of a star coorbiting with the standard thin disc as a function of the wind mass-loss rate as well as of the terminal wind velocity. { Right panel:} The analogous dependency as in the left panel, but for the case of a counter-orbiting star.}}
    \label{fig_stag_radius_star}
\end{figure*}

\begin{figure*}
    \centering
    \includegraphics[width=0.9\columnwidth]{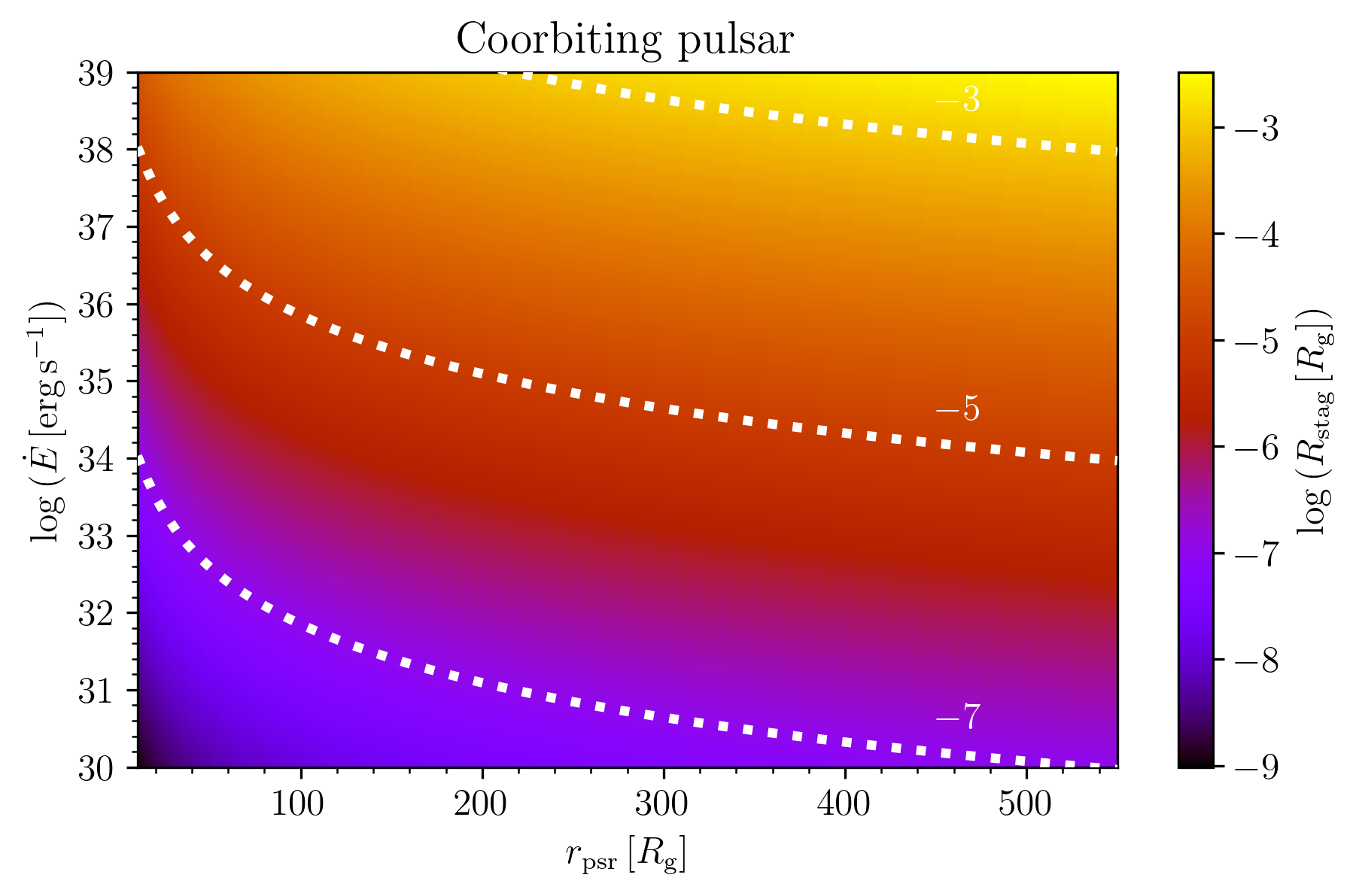}
    \includegraphics[width=0.9\columnwidth]{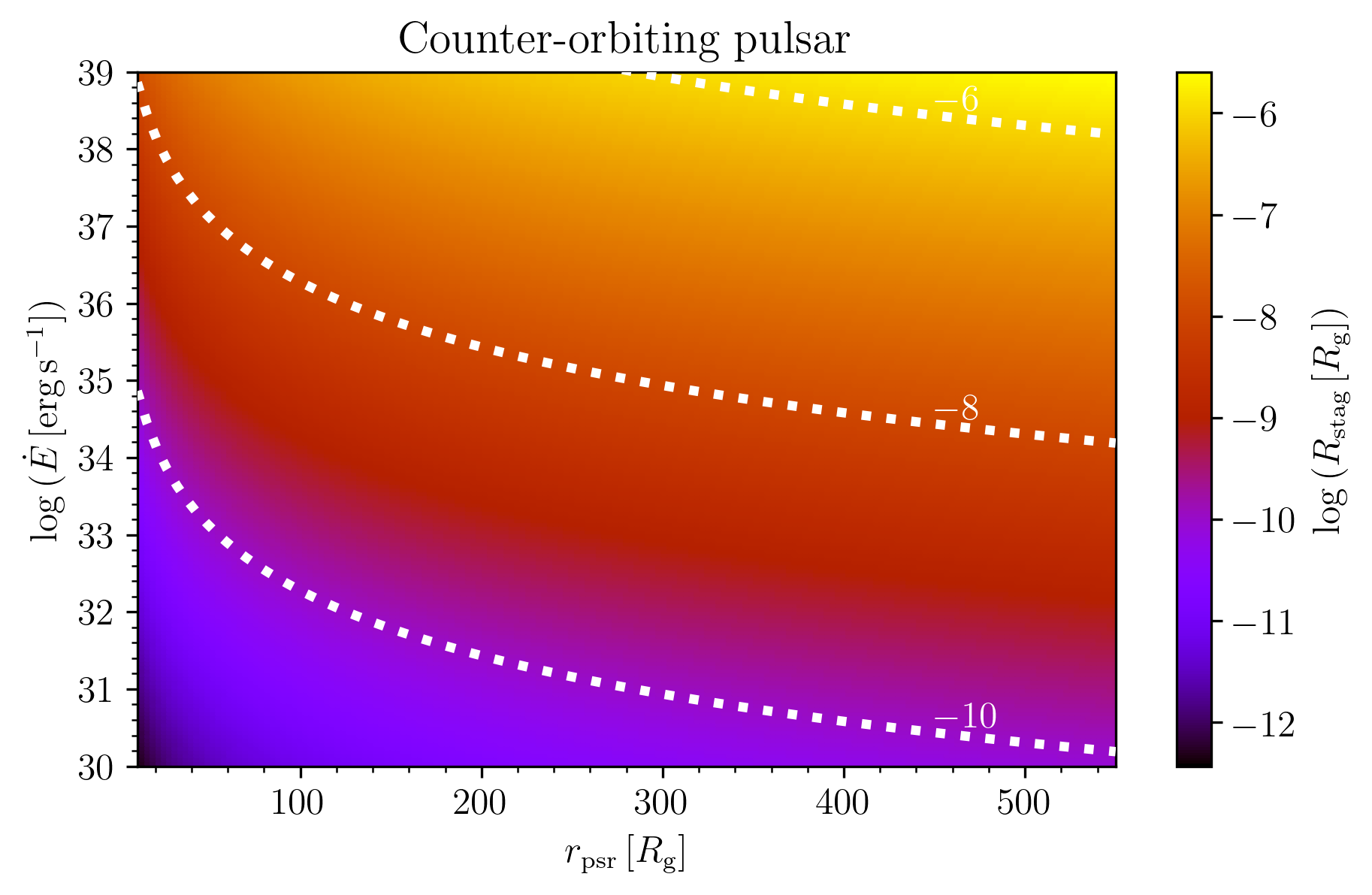}
   
    \caption{{ Stagnation radii of magnetized neutron stars -- pulsars embedded in the accretion disc surrounding $10^9\,M_{\odot}$ SMBH with $\dot{m}=0.1$ and $\alpha=0.1$. { Left panel:} Stagnation radius (expressed in gravitational radii) for the case of a pulsar coorbiting with the standard thin disc as a function of the pulsar distance from the SMBH as well as of its spin-down energy. { Right panel:} The analogous dependency as in the left panel, but for the case of a counter-orbiting pulsar.}}
    \label{fig_stag_radius_psr}
\end{figure*}

\textcolor{black}{ Here we calculate stagnation radii that correspond to wind-blowing stars and pulsars orbiting at several 100\,$R_{\rm g}$ from the SMBH, while they are embedded in the standard thin disk. Co-orbiting and counter-orbiting cases are both considered (see Figs.~\ref{fig_stag_radius_star} and \ref{fig_stag_radius_psr}). For typical parameters, the stagnation radii are found well below one $R_{\rm g}$, hence, the ability of stellar-mass objects to form detectable gaps is limited, especially in the case of standard thin disks. To create bigger gaps traceable in the accretion-disk SEDs, the mass ratio of the secondary body needs to be larger than $10^{-3}$ of the primary SMBH mass; see the main text for the related analysis and discussion.}

\bsp	
\label{lastpage}
\end{document}